\documentclass[twocolumn,preprintnumbers,nofootinbib,aps,prd,floatfix,superscriptaddress]{revtex4-2}

\usepackage{graphicx}
\usepackage{dcolumn}
\usepackage{bm}

\usepackage[left=1in, right=1in, top=1.2in, bottom=1.2in, includefoot, headheight=13.6pt]{geometry}

\usepackage{amsmath}
\usepackage{amssymb}
\usepackage{booktabs}
\usepackage[utf8]{inputenc}
\usepackage[T1]{fontenc}
\usepackage{mathptmx}
\usepackage{etoolbox}
\usepackage{float}
\usepackage[separate-uncertainty=true]{siunitx}
\usepackage{graphicx}
\usepackage{subcaption}
\usepackage{caption}
\usepackage{natbib}
\usepackage{ragged2e}
\usepackage{mathtools}
\usepackage{array}
\usepackage[table,xcdraw]{xcolor}
\usepackage{array}
\usepackage{geometry}
\usepackage{colortbl}
\usepackage{braket}

\usepackage{setspace}
\usepackage{hyperref}
\usepackage{comment}

\hypersetup{
    colorlinks=true,
    linkcolor=blue,    
    citecolor=blue,   
    urlcolor=blue       
}
\usepackage[table]{xcolor}
\usepackage{colortbl}
\newcommand{\Mpl}{M_\mathrm{Pl}}

\begin{document}

\title{Prospects for detecting new dark physics with the next generation of atomic clocks}

\author{Benjamin Elder}
\email{b.elder@imperial.ac.uk}
\affiliation{Department of Physics, Imperial College London, Prince Consort Road, SW7 2AZ, London, United Kingdom}

\author{Giorgio Mentasti}
\email{g.mentasti21@imperial.ac.uk}
\affiliation{Department of Physics, Imperial College London, Prince Consort Road, SW7 2AZ, London, United Kingdom}

\author{Elizabeth Pasatembou}
\email{e.pasatembou21@imperial.ac.uk}
\affiliation{Department of Physics, Imperial College London, Prince Consort Road, SW7 2AZ, London, United Kingdom}

\author{Charles~F.~A. Baynham}
\affiliation{Department of Physics, Imperial College London, Prince Consort Road, SW7 2AZ, London, United Kingdom}

\author{Oliver Buchmueller}
\affiliation{Department of Physics, Imperial College London, Prince Consort Road, SW7 2AZ, London, United Kingdom}
\affiliation{Department of Physics, Oxford University, South Parks Road, OX1 3PU, Oxford, United Kingdom}

\author{Carlo~R. Contaldi}
\affiliation{Department of Physics, Imperial College London, Prince Consort Road, SW7 2AZ, London, United Kingdom}

\author{Claudia de Rham}
\affiliation{Department of Physics, Imperial College London, Prince Consort Road, SW7 2AZ, London, United Kingdom}

\author{Richard Hobson}
\affiliation{Department of Physics, Imperial College London, Prince Consort Road, SW7 2AZ, London, United Kingdom}

\author{Andrew~J. Tolley}
\affiliation{Department of Physics, Imperial College London, Prince Consort Road, SW7 2AZ, London, United Kingdom}


\begin{abstract}
Wide classes of new fundamental physics theories cause apparent variations in particle mass ratios in space and time.  In theories that violate the weak equivalence principle (EP), those variations are not uniform across all particles and may be detected with atomic and molecular clock frequency comparisons. In this work we explore the potential to detect those variations with near-future clock comparisons.  We begin by searching published clock data for variations in the electron-proton mass ratio.  We then undertake a statistical analysis to model the noise in a variety of clock pairs that can be built in the near future according to the current state of the art, determining their sensitivity to various fundamental physics signals.  Those signals are then connected to constraints on fundamental physics theories that lead directly or indirectly to an effective EP-violating, including those motivated by  dark matter, dark energy, the vacuum energy problem, unification or other open questions of fundamental physics.
This work results in projections for tight new bounds on fundamental physics that could be achieved with atomic and molecular clocks within the next few years.  Our code for this work is packaged into a forecast tool that translates clock characteristics into bounds on fundamental physics.
\end{abstract}

\maketitle

\tableofcontents
\section{Introduction}
\label{sec:intro}

The world's most precise clock to date can reach relative uncertainties of $10^{-19}$ \cite{best_clock}, a level of precision which would lose less than a second over the age of the universe. This level of precision
opens up new avenues for testing fundamental physics theories by searching for variations in Nature's constants. 
In many theories beyond the Standard Model, ``fundamental constants'' such as the proton-to-electron mass ratio ($\mu$) and the fine structure constant ($\alpha$) are not fixed but can vary dynamically over time \cite{Uzan:2002vq, Uzan2011, Safronova2018}.  This behavior is exhibited in a wide variety of dark matter, dark energy, and modified gravity theories. Atomic, molecular and ion clock transition frequencies depend on those fundamental constants and are therefore sensitive to their variations~\cite{Safronova2019}.
By comparing the frequencies of two different atomic/molecular transitions, we obtain a unit-less observable that is sensitive to fundamental constant variations~\cite{Huntemann2014, clocknetwork}; the level of constraint produced by such a comparison depends on both the precision of the clocks involved and their intrinsic sensitivity to the fundamental constants under question~\cite{Ludlow2015}.

Earlier studies of temporal variations of fundamental constants searched for linear drifts in the atomic clock frequency ratios between two clock species~\cite{Fischer2004, Huntemann2014, Ludlow2015, PhysRevLett.113.210801}. Drift rates were extracted from linear fits to the data, with $\mu$ and $\alpha$ dependence derived through computational calculations of atomic structure.
More recently, frequency measurements of atomic clocks have also been used to search for oscillatory fundamental constant variations caused by hypothetical gravitational potential couplings to $\mu$ and $\alpha$ by investigating the possibility of annual sinusoidal variations caused by the varying gravitational potential as the Earth orbits the Sun. Previous studies that have searched for such variations have used neutral strontium vs.\ caesium-fountain \cite{targat2013},
hydrogen maser vs.\ caesium~\cite{Ashby2018}, and ytterbium ion vs.\ caesium~\cite{Lange2021} comparisons. These studies have searched for violations of general relativistic local position invariance with no such variations observed to date, resulting in some of the strictest constraints on fundamental theories currently known.


In parallel, over the past several decades, there have been enormous advancements in the development of different types of clocks, including atomic, molecular, and ion clocks. There are over 450 atomic clocks in over 80 national laboratories worldwide, contributing to International Atomic Time (TAI) alone~\cite{bipm2020}. It is estimated that there are thousands of atomic clocks worldwide in telecommunication networks, GPS satellites, and research institutions and universities. The most precise and accurate atomic clock has recently recorded a systematic uncertainty of $8 \times 10^{-19}$~\cite{best_clock}. Dedicated efforts to build networks of clocks with unprecedented sensitivity to temporal variations of the fundamental constants are also ongoing~\cite{qsnet,BACON_paper}.
The pace of progress in both theoretical fundamental physics and optical frequency metrology motivates a coordinated approach that integrates theory and experiment.

The purpose of this work is threefold. First, we overview wide classes of new physics theories focusing on those with specific predictions that atomic clocks can measure.  Second, we carry out a preliminary search for those signals by compiling existing publicly available data from the BIPM Circular T database~\cite{BIPMWebsite}. Third, we introduce a tool to apply statistical methods and generate constraints based on the characteristics of the clocks in question that can be used to test the sensitivity of clocks to fundamental physics theories, specifically a wide variety of modified gravity, dark energy and dark matter theories. This tool can also be used to forecast the degree to which future atomic clock advancements can result in stronger bounds on fundamental physics, and we use it to explore several such possibilities in this work.

The rest of this paper is organised as follows. Section~\ref{sec:theory_models} provides an overview of the fundamental theories that we test. Secs.~\ref{sec:forecasts} and ~\ref{sec:projected_constraints} present
an illustrative analysis using publicly available data from optical atomic clocks to produce constraints on those theories and to provide forecasts on state-of-the-art/future clock experiments and projections for their testability with atomic clocks. Section~\ref{sec:tool} presents a publicly available tool that can be used to derive constraints using the characteristics of atomic clocks as input. We conclude in Sec.~\ref{sec:conclusions}.

{\small \bf Conventions: } Throughout this work, we use natural units in which $c = \hbar = 1$, the mostly-plus metric convention, and we define the reduced Planck mass in the usual way as $\Mpl = (8 \pi G)^{-1/2} \approx 2 \times 10^{18}~\mathrm{GeV}$.

\section{Theoretical Models}
\label{sec:theory_models}

There is clear motivation from cosmology to introduce new physics.  Despite nearly a century of effort, 95\% of the matter in the universe remains unaccounted for, as does about 70\% of the total energy.  This energy appears to be driving the accelerated expansion of the universe, which can otherwise be produced by no particle in the Standard Model apart from quantum or classical vacuum energy which suffers the worse fine-tuning problem of particle physics.  Faced with this challenge, myriad of new theories have been proposed in an attempt to explain some of the aforementioned phenomena by involving either new particles, new interactions or modifications of some of the properties of the known particles of the Standard Model or the graviton.  These models  generically involve new scalar (alongside potentially higher spin) fields, either as fundamental new degrees of freedom, or (more commonly) as effective or decoupled local and four-dimensional descriptions of the new effects \cite{deRham:2023byw}.

Taking this new physics to be described by a real scalar field $\phi$, the lowest-order couplings to ordinary matter have an interaction Lagrangian that may be written as
\begin{equation}
    {\cal L}_\mathrm{int} \sim \frac{\phi}{M_p} m_p \bar p p + \frac{\phi}{M_n} m_n \bar n n + \frac{\phi}{M_e} m_e \bar e e\,.
    \label{interaction-lagrangian}
\end{equation}
Note that we have chosen to work at the level of effective couplings to protons and neutrons, which are composite particles.  
These couplings could be further decomposed into couplings to fundamental Standard Model particles~\cite{Uzan:2002vq, SevillanoMunoz:2024ayh}.  We provide an approximate mapping to one such description in Appendix~\ref{app:DM-couplings}.  Nevertheless our present approach is relatively generic and a sufficient approximation for the low-energy phenomena in atomic clocks, so \eqref{interaction-lagrangian} can be regarded as ubiquitous to  large classes of models that probe the fundamental constituents and their interactions.

In most of this work\footnote{See Appendix~\ref{app:DM-couplings} for implications and generalizations.}, we assume that nucleons couple with equal strength to the scalar field $M \equiv M_p = M_n$, so the main physical scale is captured by the differential coupling between nucleons and electrons
\begin{equation}
    M_\mathrm{eff} \equiv \left( M^{-1} - M_e^{-1} \right)^{-1}~.
\end{equation}
A non-universal coupling may appear troubling from a purely gravitational perspective, as modified gravity theories often descend from a Jordan-frame metric that depends on the scalar field and commonly results in a universal coupling to matter.  Our approach is more general, and indeed, theories with non-universal couplings like in Eq.~\eqref{interaction-lagrangian} are common targets for new physics searches, particularly ones looking for equivalence principle violation~\cite{Wagner:2012ui, Berge:2017ovy, MICROSCOPE:2022doy}.

From Eq.~\eqref{interaction-lagrangian} it is straightforward to see the dependence of the fundamental constant $\mu$ on the scalar field $\phi$. 
 Each term takes the form of a mass for its respective fermion $m_\psi(\phi) = m_\psi (1 + \phi / M_\psi)$ for $\psi \in \{p, n, e\}$. The effective mass ratio of the particles is
\begin{align} \nonumber
    \mu_\mathrm{eff}(t) &= \frac{m_p (1 + \phi(t) / M_p)}{m_e (1 + \phi(t) / M_e)} ~, \\ \nonumber
    &\approx \mu (1 + \phi(t) / M_p - \phi(t) / M_e)~, \\ 
    &= \mu \left(1 + \frac{\phi(t)}{M_\mathrm{eff}} \right)~.
    \label{mu-eff}
\end{align}
{In general the scalar field $\phi = \phi(\vec x, t)$ can vary as a function of position $\vec x$ and time $t$.  In this work we will focus on cosmological signals that are uniform in space, hence $\phi = \phi(t)$, as well as the effect of spatial variations that become time variations in Earth's reference frame as a result of Earth's orbital motion: $\phi = \phi(\vec x_\mathrm{Earth}(t), t) = \phi(t)$.}
An experiment might constrain variations in $\mu_\mathrm{eff}(t)$ over some particular timescale, that is, $\Delta \mu \equiv \mu_\mathrm{eff}(t_2) - \mu_\mathrm{eff}(t_1)$.  Dividing by the average value $\bar \mu$ we have
\begin{equation}
    \frac{\Delta \mu}{\bar \mu} = \frac{\Delta \phi}{M_\mathrm{eff}}~,
    \label{mu-variation}
\end{equation}
where $\Delta \phi$ is the change of the scalar field value over this same timescale\footnote{
An analogous procedure may be performed with the photon if the scalar is also coupled to photons, such as ${\cal L}_\mathrm{int} = \frac{1}{4} \frac{\phi}{M_\gamma} F^2$.  Assuming $\phi$ changes slowly over the time- and length-scales of the experiment, we can rescale the photon field $A_\mu \to (1 + \phi / M_\gamma)^{-1/2} A_\mu$, and we find
\begin{equation}
    \frac{\Delta \alpha}{\bar \alpha} = \frac{\Delta \phi}{M_\gamma}\,.
    \label{alpha-variation}
\end{equation}
However, in this work, we will restrict our attention to matter couplings only.
}.
This makes clear two essential ingredients for a measurable signal in atomic clocks.  First, we require a non-universal coupling to some combination of protons, neutrons, and electrons, as a universal coupling (corresponding to $M_p = M_n = M_e$, or $M_\mathrm{eff} \to \infty$) would produce $\Delta \mu = 0$.    Second, we require some variation in $\phi$ over the timescale of the experiment.  This variation depends on the specific theory in question and can be produced either locally or cosmologically.  We identify three distinct types of signals, each corresponding to a new fundamental physics scenario:
\begin{itemize}

    \item {\small \bf Modified gravity \& tests of fundamental physics:}  The inclusion of the coupling to matter in Eq.~\eqref{interaction-lagrangian} implies that matter sources the scalar field, as is the case in many models that involve a screening mechanism \cite{deRham:2023byw}, (eg. Vainshtein \cite{Vainshtein:1972sx,Babichev:2013usa,deRham:2014zqa}, Chameleon \cite{Khoury:2003aq}, or other screening mechanisms that effectively reduce the coupling between the scalar and matter \cite{Damour:1994zq,Pietroni:2005pv,Olive:2007aj,Brax:2011ja,Hinterbichler:2011ca}).  One of the strongest sources in our vicinity is the Sun, and as the Earth orbits the Sun its distance varies at the level of $\sim$1\%, producing $\phi \sim \sin (2 \pi t / \mathrm{year})$ and corresponding oscillations in $\mu$ on a one-year period, with known phase. {This is discussed in Sec.~\ref{sec:constraints-modified-gravity}.}

    \item {\small \bf Dark energy:}  Dark energy is commonly modeled as quintessence~\cite{Caldwell:1997ii, Tsujikawa:2013fta}; a scalar field rolling down some self-interaction potential $V(\phi)$.  This produces a field that varies in time $\phi = \phi(t)$, inducing corresponding variation in $\mu$.  Over the short timespans under current consideration ($\sim$ years), this variation can be reasonably approximated as linear, {$\phi \sim \sqrt{\rho_\mathrm{KE}} t$, where $\rho_\mathrm{KE} \ll (2.4~\mathrm{meV})^4$ is the amount of kinetic energy in the dark energy field. See Sec.~\ref{sec:constraints-dark-energy} for details.} 
    
    \item {\small \bf Dark matter:}  Ultra-light dark matter produces background field oscillations $\phi \sim \sin(m t)$, where $10^{-23}~\mathrm{eV} < m < \mathrm{eV}$ is the scalar particle mass \cite{Ferreira:2020fam}.  This predicts sinusoidal variations in $\mu$
    over timescales of $10^{-15}~\mathrm{s} < m^{-1} \lesssim 1~\mathrm{year}$.  {This is discussed in Sec.~\ref{sec:constraints-dark-matter}.}

\end{itemize}
The bounds from each of these signals will be compared against existing bounds on those theories, the most relevant of which are typically other clock experiments, the Planck~\cite{Planck:2018vyg} and MICROSCOPE~\cite{MICROSCOPE:2022doy} satellites, and lunar laser ranging~\cite{Nordtvedt:2003pj}.

\begin{table*}
\centering
\renewcommand{\arraystretch}{1.5} 
\arrayrulecolor{gray} 
\begin{tabular}{|>
{\centering\arraybackslash}p{3cm}|>
{\centering\arraybackslash}p{2.25cm}|>
{\centering\arraybackslash}p{2cm}|>
{\centering\arraybackslash}p{2cm}|>
{\centering\arraybackslash}p{2cm}|>
{\centering\arraybackslash}p{2.5cm}|>
{\centering\arraybackslash}p{1cm}|}
\hline
\rowcolor{gray!20} \textbf{Signal type} & \textbf{$\Delta \mu / \bar{\mu}$ signal}& \textbf{CaF/Sr (projected)} &
\textbf{Sr/Cs (projected)}&
\textbf{Circular-T~\cite{BIPMWebsite}} &
{\bf Best existing constraint} &
{\bf Units} \\ \hline

\textbf{Modified Gravity} &
$A \cos \left(\frac{2 \pi t} { \mathrm{year}}\right)$ &
$7.8\times 10^{-18}$ &
$1.4 \times 10^{-16}$ & 
$4.3 \times 10^{-17}$ &
$2.3 \times 10^{-16}$~\cite{2019Optic...6..448M} &
None \\ \hline

\textbf{Dark Energy} &
$ At$ &
$1.7\times 10^{-25}$ &
$3.1 \times 10^{-24}$ & 
$1.6 \times 10^{-20}$ &
$1.1 \times 10^{-24}$~\cite{Lange2021} &
$\mathrm{s}^{-1}$\\ \hline

\textbf{Dark Matter} &
$  \frac{A}{\omega} \cos(\omega t+\delta)$ &
$1.5\times 10^{-24}$ &
$2.8 \times 10^{-23}$ &
$3.6 \times 10^{-17}$ &
See Fig.~\ref{fig:DM-A-vs-f} &
$\mathrm{s}^{-1}$\\ \hline

\end{tabular}
\caption{Summary of types of signals in variations of $\mu$ due to fundamental physics.  All values stated are the standard uncertainties in the associated parameter $A$ as defined in the first column, i.e. $\sigma_A$.  This represents approximately the largest such signal that could reasonably escape detection.  For the dark matter signal, the values in the table show the forecast error on $A$ assuming a fiducial value of $\omega=2\pi\,\text{yr}^{-1}$ after marginalizing over that parameter as well as the phase $\delta$.   The full frequency dependence of this constraint is illustrated in Fig.~\ref{fig:DM-A-vs-f}.  {Each of these generalized signal types will be connected to fundamental physics theories in Sec.~\ref{sec:projected_constraints}.}}
\label{tab:scalar_variation}
\end{table*}

\section{Existing and simulated clock data}
\label{sec:forecasts}

In this Section we conduct a preliminary search for the signals described in Section~\ref{sec:theory_models}. We utilise publicly available data from Circular T (Section \ref{sec:circular-t}) and employ a Markov-Chain-Monte-Carlo (MCMC) and a Fisher matrix technique to project the sensitivity of different combinations of clocks (Section \ref{sec:fisher_info}). Lastly, we use simulated data from state-of-the-art atomic clocks to forecast their sensitivity to potential signals, thereby establishing constraints on relevant theoretical models based on their noise profiles. The constraint plots produced using the simulated data generation of state-of-the-art atomic clocks are presented in Sec.~\ref{sec:projected_constraints}.

Throughout this work, we use the framework presented in Ref.~\cite{clocknetwork} to relate the frequency ratios of atomic clock pairs to potential variations in the fundamental constant $\mu$. The sensitivity of a specific atomic or molecular transition, denoted by $\nu_i$, to changes in a fundamental constant represented by $X = \{\alpha, \mu\}$, is quantified by a sensitivity coefficient denoted as $K_X$.  

The sensitivity to variations in a particular fundamental constant $X$ for a given frequency ratio R = $\nu_1$ / $\nu_2$ is directly proportional to the difference between the sensitivity coefficients:
\begin{equation}
\frac{\Delta R}{R} = \left[K_{X,1} - K_{X,2} \right]\frac{\Delta X}{X}\,.
\label{eq:sensitivity}
\end{equation}
A higher value of $K_X$ indicates greater sensitivity of the specific transition to variations of X. For optical electronic transitions $K_{\mu, opt} = 0$; for molecular vibrational transitions $K_{\mu, vib} = - \frac{1}{2}$; and for hyperfine microwave (MW) transitions $K_{\mu, MW} = - 1$~\cite{qsnet, Sherrill_2023}. For $K_\alpha$, the sensitivity coefficients depend on the detailed atomic structure of the atom or molecule considered. See Ref.~\cite{clocknetwork} for a discussion on how these sensitivity coefficients are derived.

The transition frequencies of the atomic clocks considered in this work can be written in the form~\cite{clocknetwork}:
\begin{align} \nonumber
v_{\mathrm{opt}} &=A \cdot F_{\mathrm{opt}}(\alpha) \cdot c R_{\infty}\,, \\
v_{\mathrm{vib}} &=C \cdot \left(\frac{m_e}{m_p}\right)^{\frac{1}{2}} \cdot c R_{\infty}\,, \nonumber \\
v_{\mathrm{MW}} &=B \cdot \alpha^2 F_{\mathrm{MW}}(\alpha) \cdot \frac{m_e}{m_p} \cdot c R_{\infty}\,,
\end{align}
with the Rydberg constant $R_{\infty}=\frac{c}{4 \pi \hbar}  \alpha^2 m_e$. The $F(\alpha)$ terms account for relativistic perturbations to the atomic structure, which can be significant in heavier elements. 

\subsection{Testing the models on Circular T data}
\label{sec:circular-t}

To provide a baseline for constraints on the classes of signal described in this paper, we performed a preliminary study using publicly available data published monthly by the BIPM through Circular T~\cite{BIPMWebsite}. The BIPM provides a monthly release of the BIPM's Time Department data, allowing for local realisations of UTC(k), which are maintained by national institutes, to trace back to the Coordinated Universal Time (UTC). UTC is a globally recognised time standard used as a reference for timekeeping and is established using data from atomic clocks operated by over 80 contributing institutes worldwide~\cite{UTC1, UTC2}. Since the definition of the SI second is based on the frequency of the hyperfine transition in caesium, most clocks contributing to TAI are caesium-based with $K_\mu = -1$. However, in recent years, contributions from clocks based on optical transitions have emerged~\cite{BIPMWebsite} which have $K_\mu=0$. In this work, we exploit this difference in sensitivity to test for variation in $\mu$ by assuming that the mean value of TAI is entirely determined by the caesium transition, and comparing individual strontium and ytterbium optical clocks against this total mean to extract a cross-species frequency ratio. 

The data used to produce the constraints and the mapping from that data to $\Delta \mu / \mu$ are presented in Appendix \ref{app:circular-t}. For this analysis, values of $d$, the negative fractional frequency deviation of TAI, for individual optical frequency clocks as reported in Circular T are used as described in the same Appendix.
It is assumed that this value is proportional to $ \Delta R / R$ and therefore to $\Delta \mu / \mu$ where $R$ is the ratio of frequencies of the optical clock considered to the mean caesium transition frequency.

After collating the data for all the optical clocks contributing to Circular T (Fig.~\ref{fig:dvalues}, in Appendix \ref{app:circular-t}), we use Bayesian inference to evaluate the posterior distribution of the unknown parameters in our models. Three models were used: (a) a sinusoidal model with a fixed period of one year with the amplitude as the only free parameter, (b) a linear model with the gradient as the unknown parameter, and (c) a sinusoidal model with an unknown period and phase. We use the {\tt emcee} Python package \cite{emcee} to perform an MCMC analysis of the (log)-posterior distribution in the theoretical parameters given the data (see Appendix \ref{bayesian_inf}).

The constraints produced from this analysis are presented in Table \ref{tab:scalar_variation} for comparison with the other constraints on the associated parameters stated. It should be noted that this analysis is limited by assuming that the quoted uncertainties are reliable and that there are no correlations between measurements of the same clocks.  

We also assume that all optical clocks from all the different laboratories are affected in the same way by the physical phenomena, and that the uncertainties in the measurement of $\frac{dR}{R}$ is affected by a gaussian noise, with no correlations between different clock pairs.
As a point of comparison, the maximum allowed signals in the data that could have escaped detection are summarised in Table~\ref{tab:scalar_variation}. 

\subsection{Forecasts with state-of-the-art atomic clocks (Fisher information)}
\label{sec:fisher_info}

To forecast the possible observation of variations in fundamental constants using current state-of-the-art atomic clocks, a Fisher forecast method is adopted, along with a MCMC analysis, to evaluate the projected sensitivity of the experiments to potential signals of fundamental constant variations. Those methods are widely used in cosmology~\cite{2020_cosmo_book} as tools for parameter estimation forecasts, which inform the development of future experiments.

The Fisher matrix method is a Bayesian inference approach, and the method is used to detect potential variations in the ratio of the measured frequencies of two clocks over time.
For the clock projections, we simulate a datastream sampled with a cadence of 1 second, coming from the superposition of the frequency ratio $\Delta R/R$ induced by our set of theories and the instrumental noise from a realistic noise model built on the physical properties of the clocks considered.
The mathematical framework of this technique is presented in Appendix \ref{sec:fisher}. This data stream is Fourier transformed into the frequency domain to facilitate the analysis of the signal characteristics and separate it from noise. The noise is assumed to be Gaussian and stationary with zero mean, and we assume the datastream is ungapped. The relaxations of these assumptions are not expected to significantly impact the forecast, even if the analysis will become much more challenging in practice. Some details regarding this point are given in Appendix \ref{sec:fisher}.
The Fisher approximation of the posterior distribution of the unknown parameters is then analytically computed to have a quick and reliable estimate of the confidence intervals to produce analytical and interpretable forecasts. An MCMC evaluation of the full forecast posterior distribution has been done to check the validity of the Fisher approximation.  The clock characteristics used in this part of the analysis can be found in Appendix~\ref{app:current_clocks} Table \ref{tab:clock_stab_acc}.

\subsection{Forecasts - Simulated Data}
\label{sec:simulated_data}

\begin{figure}[t]
    \centering
    \includegraphics[width=\linewidth]{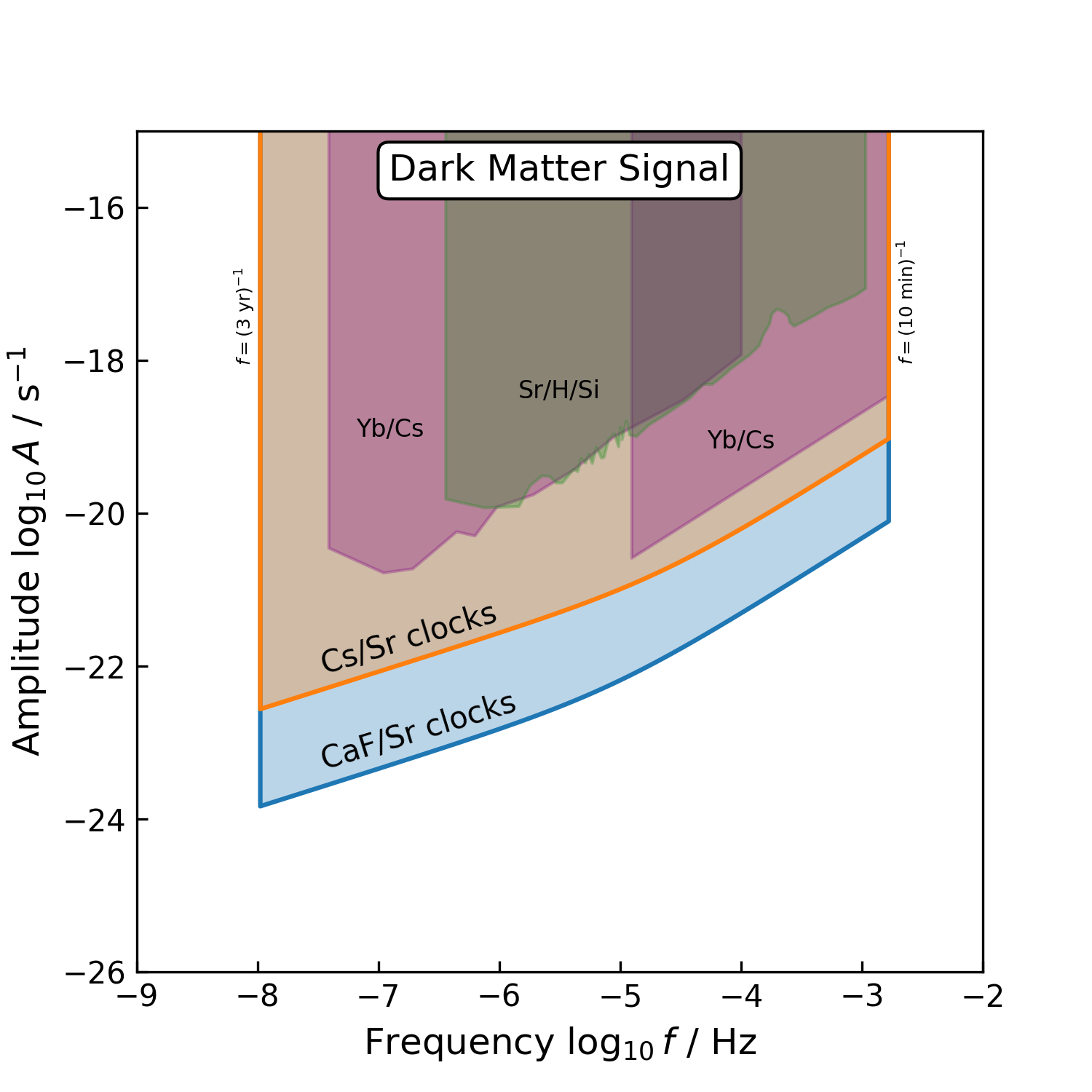}
    \caption{\small Comparison of projected and existing constraints on oscillations in $\mu_\mathrm{eff}(t) / \bar \mu = 1 + A / \omega \cos (\omega t + \delta)$.  This signal is associated with ultralight dark matter theories.  Also shown are currently-leading constraints from clocks in this frequency range~\cite{Kennedy:2020bac, Kobayashi_2022, Sherrill_2023}.  {Note that this figure is in terms of the ordinary frequency $f$, not the angular frequency $\omega = 2 \pi f$.}}
    \label{fig:DM-A-vs-f}
\end{figure}

This Section presents the method used to create simulated clock data to derive constraints on the various models using Bayesian inference. A numerical simulation is implemented to generate synthetic datasets consisting of a signal embedded in noise.

The analysis is carried out in the frequency domain instead of the time domain. This is because the noise profile includes a {$1/f = 2 \pi / \omega$} dependence noise term (flicker/pink noise), which implies correlations in the noise of the data points in the time domain, making the analysis more complex and computationally expensive as the correlations would need to be explicitly handled. Working in the frequency domain, where the covariance matrix of the data is diagonal, removes this requirement and simplifies the analysis.

We first generate noise with white and pink noise components for two clocks by modelling the noise profile of the clocks using the one-sided power spectral density (PSD) for a single clock, as indicated in Eq.~\ref{PSD_equation}. The values for the coefficients used for each clock are presented in Table \ref{tab:clock_stab_acc}. The total duration of the simulated signal is three years with a cadence of one second. Two clocks are required for sensitivity to variations in fundamental parameters, therefore, the combined resulting PSD of the two clocks is the combination of the individual PSDs for each clock assuming that the noise streams of the individual clocks are uncorrelated. The resulting PSD is a summation of the individual clocks PSD. The combinations of clocks used are presented in Fig.~\ref{fig:psd_sensitivity} (bottom plot).

The signal predicted by the theories considered is parametrized in the frequency domain and then added to the simulated noise stream. The injected signal is determined in the frequency domain for all different scenarios by taking the Fourier transform of the time-domain signals outlined in Table~\ref{tab:scalar_variation}. The MCMC analysis is then performed on the resulting dataset to obtain samples from the posterior distribution of the unknown parameters.

The values for the coefficients of the PSD are included in Table~\ref{tab:clock_stab_acc}. The MCMC sampling method described in the previous section is used to derive the posterior distributions for the different free parameters in question for the different models. The results are then used to investigate the relationship of different parameters of the models in question. Plots investigating the relationship of different parameters are presented, and the constraints on the space parameters imposed by different pairs of atomic clocks are also shown.

This was done for each of the signals listed in Table~\ref{tab:scalar_variation}.  The results for two representative samples of clock combinations are also listed in that table.  The ``dark matter'' signal is slightly more involved because the threshold for detection of $A$ now depends on the typical frequency of the signal $\omega$.  We therefore include additional information for this case as a worked example.  The approach may be generalized straightforwardly for the other two fundamental physics signals types. One key distinction of the modified gravity signal is that the phase of the oscillations is set by the position of the Earth relative to the Sun, {the details of which are discussed in Sec.~\ref{sec:constraints-modified-gravity}.}

For the dark-matter-like signal,
we show the upper bound that the experiment can pose on the amplitude parameter A as a function of the value of the frequency of the signal, which is the curve
\begin{align}
A > \sigma_A= \sigma_A(\omega_0)\frac{\omega}{\omega_0}\frac{N(\omega)}{N(\omega_0)}
\end{align}
where $N(\omega)$ is the noise power spectral density at frequency $\frac{\omega}{2\pi}$ and $\sigma_A(\omega_0)$ is the value of the forecast error in the measurement of the parameter $A$ at the reference frequency $\omega_0=\frac{2\pi}{1\rm yr}$. The curve is shown in Fig.~\ref{fig:DM-A-vs-f}, with the values of $\sigma_A(\omega_0)$ displayed in Table \ref{tab:scalar_variation}, and the derivation is detailed in Appendix \ref{sec:fisher}.

\subsection{Existing varying-$\mu$ searches}
Variations in $\mu$ have been previously discussed in the literature in a variety of contexts.  In this section, we briefly review a few of the most relevant existing measurements.

Six years of clock data were examined for oscillatory signals associated with dark matter with an Rb/Cs clock pair~\cite{Hees:2016gop}.  However, this particular clock pair has a sensitivity coefficient difference for $\mu$ of 0, making it difficult to perform a direct comparison in this work.  Indeed, a nonzero sensitivity coefficient $K_\mu$ (described at the beginning of Sec.~\ref{sec:forecasts}) is a key advantage of the clock pairs we focus on in this work.  In the interest of facilitating comparison, we translate our bounds into the dark matter model used in that work in Appendix~\ref{app:DM-couplings}.

A comparison of Cs/Yb/Sr clocks was combined to produce estimates of signals equivalent to our modified gravity and dark energy signals~\cite{2019Optic...6..448M}.  The uncertainty on the amplitude of the yearly modified gravity signal was $\sigma_A = 2.3 \times 10^{-16}$, which is directly comparable to our results.  It is seen in Table~\ref{tab:scalar_variation} that this is roughly in line with our Sr/Cs clock projections and about 1.5 orders of magnitude weaker than our CaF/Sr projections.  A comparison of non-continuous clock data over a period of roughly 13 years was also undertaken to search for a linear drift in $\mu$, which was measured to be $\dot \mu / \bar \mu = (5.3 \pm 6.5) \times 10^{-17} / \mathrm{yr}$.  Converting this uncertainty into units of $s^{-1}$ allows for a direct comparison to our dark energy signal, {and rules out $A \gtrsim 2.1 \times 10^{-24}~\mathrm{s}^{-1}$.
Referring to Table~\ref{tab:scalar_variation},} we see that the measurement is roughly in line with our Sr/Cs clocks projection and about one order of magnitude worse than the CaF/Sr projection.

Frequency comparisons between a strontium clock, silicon cavity, and hydrogen maser were performed to search for ultralight dark matter~\cite{Kennedy:2020bac}.  Those bounds are stated in terms of a scalar-electron coupling parameter $d_{m_e}$ which is related to our $M_\mathrm{eff}$ by
\begin{equation}
    M_\mathrm{eff} = \frac{\sqrt 2 \Mpl}{d_{m_e}}~.
\end{equation}
Those bounds are included in Fig.~\ref{fig:dark-matter}.  To translate the bounds to the amplitude signal of Fig.~\ref{fig:DM-A-vs-f}, we use Eq.~\eqref{dm-mu-variation} to compute the signal amplitude as
\begin{equation}
    A = \frac{\sqrt{2 \rho_\mathrm{DM}}}{M_\mathrm{eff}}
\end{equation}
where we have used the local density for dark matter as described in Sec.~\ref{sec:projected_constraints}.
Also of note is a search using Yb/Cs clocks which ran for 298 days~\cite{Kobayashi_2022}, and is also included in Fig.~\ref{fig:DM-A-vs-f}.  In a similar vein there are a number of experiments that constrain ultralight dark matter in this range, particularly NANOGrav~\cite{NANOGrav:2023hvm} which appears on Fig.~\ref{fig:dark-matter}.  More generally a number of constraints on the ultralight dark matter-Standard Model interactions are summarised in~\cite{AxionLimits}.

More recently, data from Yb, Sr, and Cs clocks was used to search for oscillatory ultralight dark matter signals~\cite{Sherrill_2023}.  This data is in 600s intervals and was collected over a period of two weeks.  Since the dark matter signal in Table~\ref{tab:scalar_variation} is normalized to 1 year, this result is omitted from the table, but it is included in Figs.~\ref{fig:DM-A-vs-f} and \ref{fig:dark-matter}.  These bounds cover a similar frequency range as explored here, and are approximately one order of magnitude weaker than our current projections.

The coupling between clock frequencies and the Newtonian gravitational potential were measured in~\cite{Ashby2018}, and was summarized for a variety of atomic species.  Unfortunately, this result is not directly comparable to our approach without additional theoretical modelling, as the modified gravity models we focus on in this work fall off differently from the Newtonian potential.  In future work, it would be interesting to translate their results into bounds on specific modified gravity models.  A similar approach was taken in~\cite{Lange2021}, although that work also constrained a linear temporal variation in $\mu$ to be $\dot \mu / \mu = (-8 \pm 36) \times 10^{-18}~ \mathrm{yr}^{-1}$.  This corresponds to a maximum dark energy signal amplitude of $A \leq 1.1 \times 10^{-24} ~\mathrm{s}^{-1}$, making it, along with MICROSCOPE, one of the best current constraints on the EP-violating dark energy signal.

It was recently proposed to search for chameleon particles by employing two identical clocks, where one of the clocks is placed inside a massive source object (or with such an object nearby) to search for redshift effects~\cite{Levy:2024vyd}.  Such effects could, in principle, be searched for with the clocks under present consideration as well, although in this work, we focus on equivalence-principle violation searches through varying-$\mu$ as this is a unique feature of the clock combinations we are considering.  It would be interesting to explore the potential for these clocks to search for the above signal, particularly for short-range forces, in future work.

\section{Projected constraints on fundamental physics}
\label{sec:projected_constraints}

This Section presents projected constraints on fundamental physics theories from state-of-the-art clocks for all three signals described in Section \ref{sec:theory_models}.  These signals are associated with modified gravity, dark energy, and dark matter, respectively.  In each case, the underlying physics is modelled as a single real scalar field $\phi$, coupled to matter via the interaction Lagrangian given in Eq.~\eqref{interaction-lagrangian}.  The coupling to matter, along with spatial and/or temporal variations in the local scalar field value, leads to variations in the apparent electron-proton mass ratio $\mu_\mathrm{eff}$.  In this section, we mainly emphasize the results on a few of the most novel or promising routes for detecting new fundamental physics rather than a broad overview of all possible theories that could be tested.

\subsection{Fifth forces and screening}
\label{sec:constraints-modified-gravity}

A direct scalar-matter coupling of the form Eq.~\eqref{interaction-lagrangian} generically implies that the scalar field mediates a new ``fifth'' force between matter particles.  There are a wide range of fifth force models in the literature, which are reviewed in \cite{Joyce:2014kja, CANTATA:2021asi, Brax:2021wcv, deRham:2023byw}.  While clocks are not directly sensitive to this new force, they are sensitive to that force's potential, which is proportional to the local field value $\phi$, as is clear from Eq.~\eqref{mu-eff}.  The Sun, like all massive bodies, is a source for that potential.  As the Earth orbits the Sun, the Earth-Sun distance varies at the level of $\sim 1\%$, leading to variations  in $\phi$ (and hence the proton-electron mass ratio $\mu$) on a one-year period.  A similar approach was taken in \cite{PhysRevD.106.095031}, although that work focused on theories without scalar self-interactions and on variations in the fine structure constant $\alpha$ and particle masses.

For our purposes, many models can be analyzed via the same procedure where the additional degrees of freedom behave as a scalar in some limit. Then we can first consider the scalar field's equation of motion around the Sun in order to find $\phi(r)$, where $r$ is the distance to the Sun.  Then, we supply $r(t)$, which describes the Earth's distance to the Sun as a function of time.  This allows us to compute variations in $\mu_\mathrm{eff}$ as a function of time via Eqs.~\eqref{mu-eff}.  In this Section, we focus on a pair of modified gravity models, and further models are discussed in Appendix~\ref{app:modified_gravity}.

Leading theories of gravity which depart from four-dimensional General Relativity on large cosmological scales are associated with theories where the graviton is effectively  massive either as a resonance or as an effect from extra dimensions \cite{Dvali:2000hr,Dvali:2000rv,deRham:2007xp,deRham:2007rw} or as a four-dimensional local theory of gravity~\cite{deRham:2010kj,deRham:2010tw,deRham:2014zqa}.  While Lorentz-invariant theories of massive gravity (whether they are effective or not) involve additional degrees of freedom (or polarizations),  that could in principle mediate fifth forces, they also come hand in hand with a Vainshtein screening mechanism \cite{Vainshtein:1972sx,Deffayet:2001uk} which is best understood in the decoupling limit \cite{Luty:2003vm}, where one of the degrees of freedom behaves as a scalar field (dubbed the Galileon) \cite{Nicolis:2008in,deRham:2010eu}. The Vainshtein mechanism naturally ensures that scalar self-interactions suppress the fifth force sourced by massive objects and lead to a smooth massless limit. 
This theory is introduced in more detail in Appendix~\ref{app:Galileon}, while here we give a brief overview of this theory's relevant properties.  

The Galileon is characterised by both a scalar-matter interaction set by the parameter $M$, as well as several nonlinear self-interaction terms which are set by the parameters $\Lambda, c_4$.  {Note that $\Lambda$ is related to the scale of the graviton mass $m_g$ in this theory by \cite{deRham:2016nuf}
\begin{equation}
    \Lambda^3 \approx m_g^2 \Mpl~.
    \label{graviton-mass-scale}
\end{equation}
}
Around a spherical source object of mass $m_\mathrm{obj}$ the Galileon's equation of motion is:
\begin{equation}
    \left( \frac{\phi'}{r} \right) + \frac{2}{\Lambda^3}\left( \frac{\phi'}{r} \right)^2 + \frac{2 c_4}{\Lambda^6}\left( \frac{\phi'}{r} \right)^3 = \frac{m_\mathrm{obj}}{4 \pi M r^3}~.
    \label{Galileon-eom}
\end{equation}
The terms on the left follow from the theory's ordinary kinetic term, its cubic self-interactions, and its quartic self-interactions in its Lagrangian, respectively.
There are three parameters: $\Lambda$ sets the overall strength of the Galileon's self-interactions, $c_4$ controls the relative strength of the quartic interactions, and $M$ controls the strength of the coupling to matter.  Note that in the interest of simplicity we have coupled the Galileon to nucleons with equal strength and not at all to electrons (that is, in this section we take $M_e \to \infty$)\footnote{Strictly speaking, the source in Eq.~\eqref{Galileon-eom} should not be the total mass, but only the fraction of the source's mass made up of nucleons.  However, this distinction corrects the value of $m_\mathrm{obj}$ at the level of a part in $10^3$ which makes a negligible difference to our present estimates.}.

Far away from the source, the cubic and quartic terms are strongly suppressed relative to the kinetic term, which may be intuitively understood by observing the additional factors of $\phi' / r$ in the equation of motion.  As such, at large $r$, the field's behaviour is dominated by the ordinary kinetic term, giving a field profile that goes as $\phi \sim 1/r$, much like in Newtonian gravity.  Closer to the body, the Galileon's self-interactions become significant, and the cubic interactions dominate the left-hand side of Eq.~\eqref{Galileon-eom}.  Within this region, the falloff is reduced to $\phi \sim 1/\sqrt{r}$.  The boundary between these two regimes is termed the ``cubic Vainshtein radius'' $R_3$.  Closer still to the massive body, the quartic terms dominate, and the field profile tends towards a constant.
Summarizing, we have
\begin{equation}
    \phi(r) \sim \left\{\begin{array}{lcc}
        m_\mathrm{obj}^{1/3} &  {\rm for} \quad & \quad\quad\quad  r < R_4^4 / R_3^3 ~, \\
        m_\mathrm{obj}^{1/2} r^{-1/2} & {\rm for} \quad & R_4^4 / R_3^3  < r < R_3~, \\
        m_\mathrm{obj} r^{-1} & {\rm for} \quad &  R_3 < r~.
    \end{array}\right.
\end{equation}
The Vainshtein radii $R_{3,4}$ depend on the mass of the source and parameters of the theory and are given in Appendix~\ref{app:Galileon}.

As the Earth orbits the Sun, its distance varies as
\begin{align} \nonumber
    r(t) &= a \frac{1 - \epsilon^2}{\epsilon \cos (2 \pi t / \mathrm{year}) + 1}~, \\
    &= a \left(1 - \epsilon \cos (2 \pi t / \mathrm{year}) + O(\epsilon^2) \right)~,
    \label{earth-sun-distance}
\end{align}
where $a = \mathrm{AU}$ is the average Earth-Sun distance and $\epsilon = 0.0167$ is the eccentricity of Earth's orbit. 
This leads to yearly sinusoidal variations in $\phi$, and therefore also in $\mu$ via Eq.~\eqref{mu-eff}.  These variations are to be compared with the result of the statistical analysis searching for sinusoidal variations in $\mu$ with a one-year period, which is summarised in Table~\ref{tab:scalar_variation}.  Note that (unlike the case with dark matter) the phase of this signal is known and is set by the position of the Earth relative to the Sun.

To solve for the scalar field, we Taylor expand as
\begin{equation}
    \phi(r) = \phi(a) + \phi'(a) (r - a)~.
\end{equation}
Only the second term varies in time via $r = r(t)$.  Combining, we find
\begin{equation}
    \frac{\mu(t)}{\bar \mu} = 1 +  \frac{\phi'(a) a \epsilon}{M} \cos(2 \pi t / \mathrm{year})~.
\end{equation}

\begin{figure}[H]
    \centering
    \includegraphics[width=\linewidth]{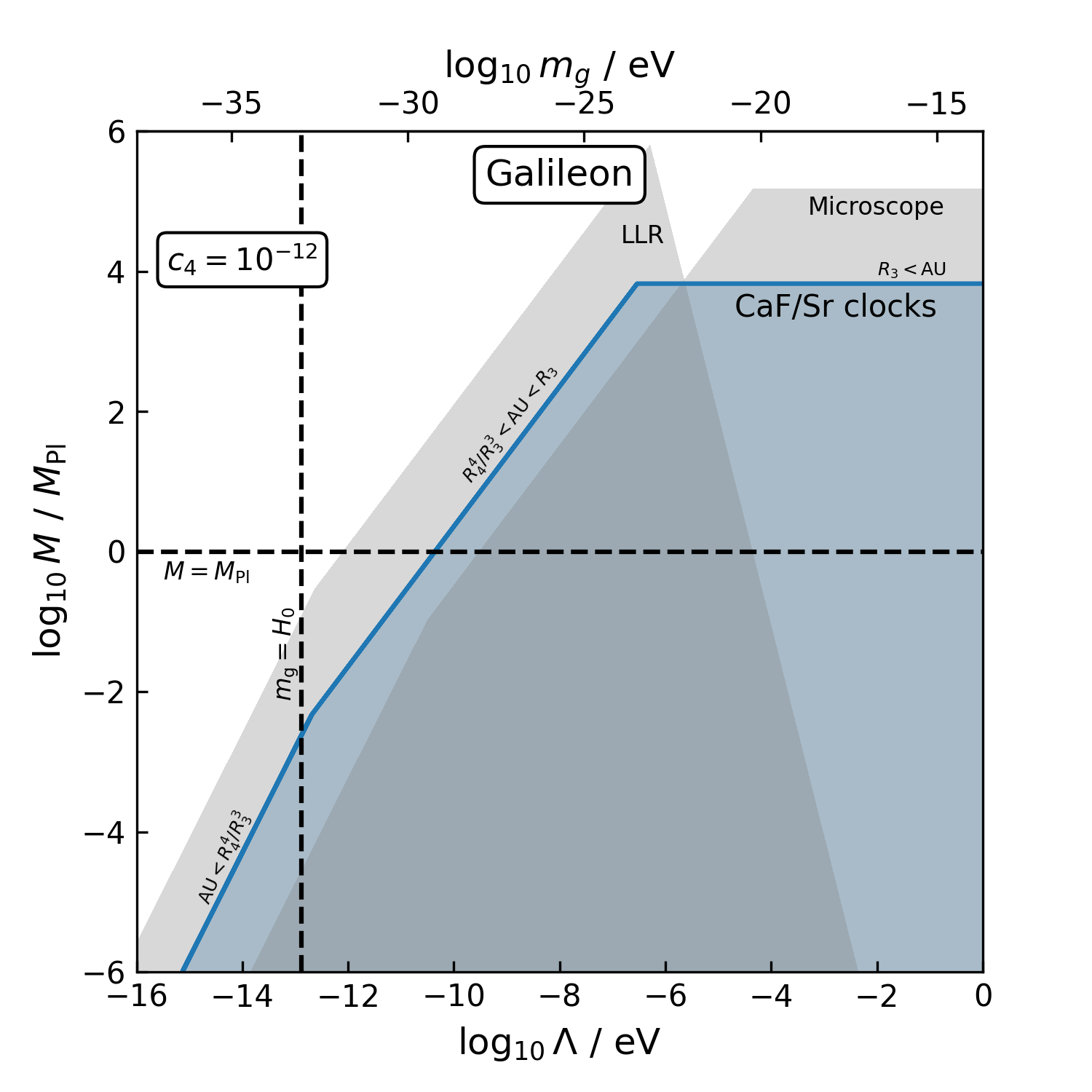}
    \caption{\raggedright Constraints on Galileon parameter space from a CaF/Sr clock pair over a period of three years.  The different power laws in the curve correspond to regions where quadratic, cubic, and quartic terms of Eq.~\eqref{Galileon-eom} each dominate.  The gray regions are ruled out by the MICROSCOPE experiment and lunar laser ranging, which are discussed in Appendix~\ref{app:Galileon}.  Also indicated are $M = M_\mathrm{Pl}$, which corresponds to a gravitational-strength matter-scalar coupling, and the $\Lambda$ scale that corresponds to a graviton mass proportional to the current Hubble scale, as given by Eq.~\eqref{graviton-mass-scale}.}
    \label{fig:MG-Galileon}
\end{figure}

The result of this analysis is shown in Fig.~\ref{fig:MG-Galileon}.  This analysis was done for a CaF/Sr clock pair over a period of three years, which was shown in Table~~\ref{tab:scalar_variation} to be one of the most sensitive clock systems to the yearly periodic modified gravity signal.

We can make some general observations about these results.  First, we can identify the three distinct regimes, depending on the size of $\Lambda$, corresponding to the regions where the quadratic, cubic, and quartic self-interaction terms each dominate.  When $\Lambda$ is large, the Vainshtein radii are smaller than the Earth-Sun distance and the Sun is therefore unscreened.  Decreasing $\Lambda$, we see a region dominated by the cubic self-interaction term, and constraints weaken because of the stronger screening.  Decreasing $\Lambda$ further still we see a region dominated by the quartic term, with constraints that are weaker still thanks to the strong screening effect.

Also plotted on the figure are the constraints from MICROSCOPE~\cite{Berge:2017ovy,MICROSCOPE:2022doy}, a satellite-borne test of the universality of free-fall towards Earth.  These bounds are discussed in greater detail in Appendix~\ref{sec:microscope}.  When screening is inactive (that is, at large $\Lambda$), the bounds from MICROSCOPE are superior to the projected bounds from clocks.  However, in the screened regime the situation reverses and clocks appear more promising.  This may be understood intuitively by examining Eq.~\eqref{Galileon-eom}.  In the unscreened regime, which is to say when $R_\mathrm{V3,4}$ are both smaller than the Earth-Sun distance and Earth-MICROSCOPE distance, the signal is proportional to $m / r$.  For our clock setup we take $m = m_\odot$ and $r = \mathrm{AU}$, while MICROSCOPE has the much smaller $m = m_\oplus$ and $r = R_\mathrm{microscope} \approx 7000~\mathrm{km}$.  MICROSCOPE measures a much smaller source (the Earth) but at a far shorter distance, which more than compensates for the smaller source.  Hence the constraints from MICROSCOPE are comparatively stronger.  However, once screening becomes active, the signal instead goes as $m^{1/2} / \sqrt{r}$ and $m^{1/3}$ for the cubic and quartic regimes, respectively.  In these cases, the falloff of the signal is smaller, so there is a comparatively smaller penalty for the Sun-clock measurement's longer baseline.  It is for this reason that the clock constraints compare favorably against MICROSCOPE within the screened regime of the Galileon.  Notably, we find that the constraint on $M$ is improved by an order of magnitude over MICROSCOPE for the $\Lambda$ that gives a graviton mass of order the Hubble scale, $m_g \sim H_0 \sim 10^{-33}~\mathrm{eV}$.

\begin{figure}[h!]
    \centering
    \includegraphics[width=\linewidth]{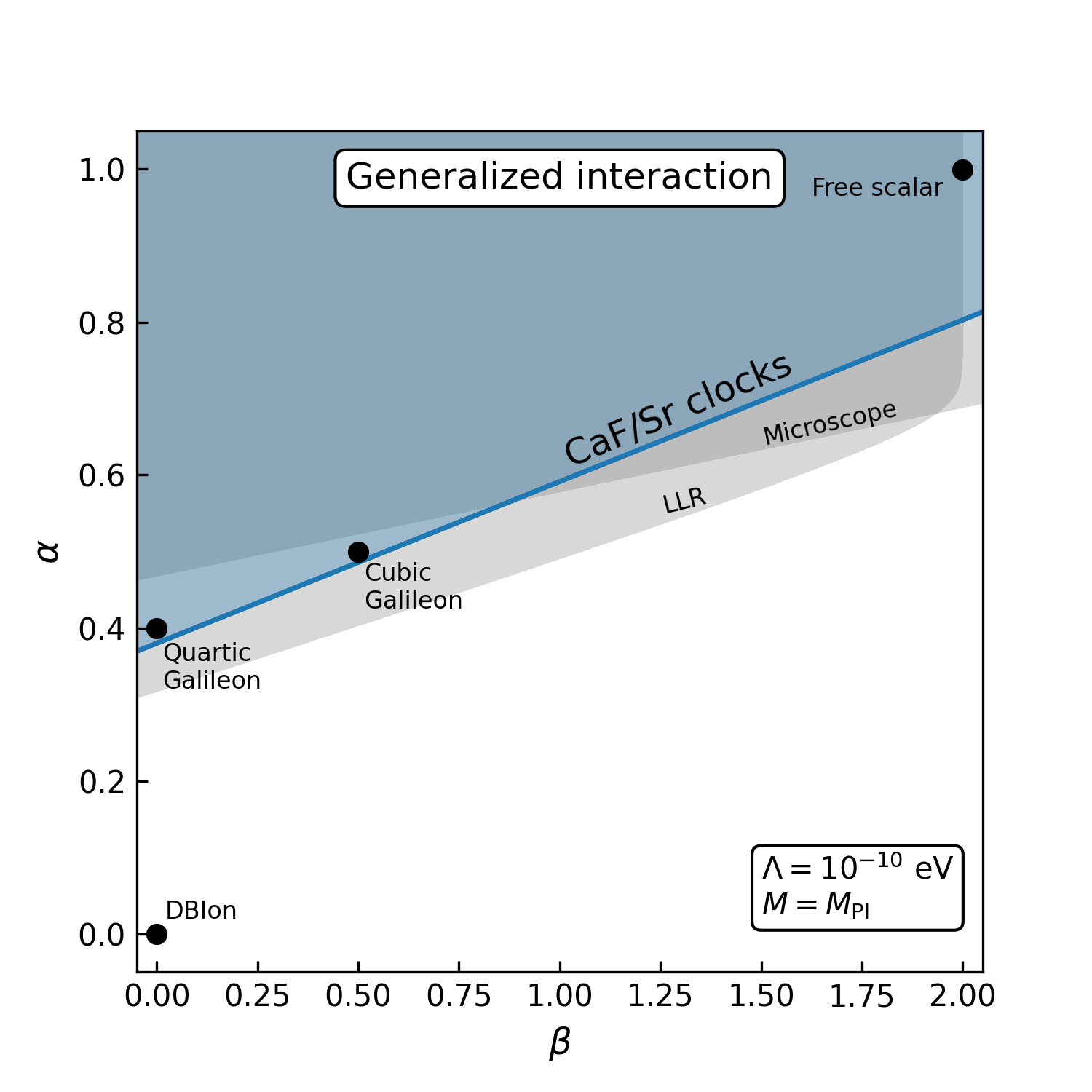}
    \caption{\raggedright Upper panel: constraints on the space of parameters $(\Lambda,M)$ in the generalized interaction modified gravity model \eqref{generalized-profile}. The lines define the lower bounds in the region of parameters $(\Lambda,M)$ that can be ruled out by a CaF/Sr clock pair over a three-year observation time.  Particular theories are highlighted: the free scalar, cubic Galileon, and quartic Galileon corresponding to each of the three regimes identified in Eq.~\eqref{Galileon-eom}, and also generalized ones that include DBI~\cite{deRham:2010eu}.}
    \label{fig:MG_gen_int}
\end{figure}

This pattern is also evident in our second sample theory.  For this, we present a parametrized interaction model.  This model describes the field sourced by a spherical object of mass $m_\mathrm{obj}$ as
\begin{equation}
    \phi' = \Lambda^2 \left( \frac{m_\mathrm{obj}}{8 \pi M} \right)^\alpha \left(\Lambda r\right)^{-\beta}~.
\end{equation}
Although we did not derive this from a Lagrangian, this phenomenological description encapsulates the behavior of a large number of modified gravity theories, which correspond to different choices of the parameters $\Lambda, M, \alpha,$ and $\beta$.
Once again, for our projected clock constraints, we will choose $m_\mathrm{obj} = m_\odot$ and $r = r(t)$ is the Earth-Sun distance over time given by Eq.~\eqref{earth-sun-distance}.  The scalar field parameters $\alpha, M$ characterize the scalar-matter coupling, $\Lambda$ is a mass scale related to the scalar field's self-couplings, and $\beta$ describes how quickly the scalar field's potential falls off with distance from a source which is affected by the nature of the interactions (see for instance \cite{deRham:2007rw} for an example where additional extra dimensions lead to a different scaling).  As a simple example, a free scalar field with a gravitational-strength coupling to matter corresponds to $\alpha = 1, \beta = 2, M = M_\mathrm{Pl}$, and $\Lambda$ cancels out.
Projected clock constraints on this generalised interaction model for several sets of fiducial parameter values are presented in Fig.~\ref{fig:MG_gen_int}.  Once again, we can observe the same general pattern as the Galileon: theories where the field drops off more slowly than $1/r$ compare more favorably against MICROSCOPE.  For both modified gravity models, we see that the projected clock bounds are currently weaker than those deriving from lunar laser ranging (LLR)~\cite{Nordtvedt:2003pj, Dvali:2002vf, Brax:2011sv, Tsujikawa:2019pih}.  A further factor of 100 improvement in clock sensitivity beyond what is projected here would result in clock bounds that are competitive with LLR.

\subsection{Dark energy}
\label{sec:constraints-dark-energy}

One of the simplest models of dynamical dark energy is quintessence~\cite{Caldwell:1997ii, Tsujikawa:2013fta}, which involves a scalar field field rolling down some self-interaction potential $V(\phi)$:
\begin{equation}
    {\cal L}_\phi = - \frac{1}{2} (\partial \phi)^2 - V(\phi)~.
\end{equation}
On cosmological scales, the field depends only on time, $\phi = \phi(t)$ and is, therefore, a perfect fluid with an equation of state and density
\begin{align} \nonumber
    w &= \frac{ \frac 1 2 \dot \phi^2 - V(\phi)}{ \frac 1 2 \dot \phi^2 + V(\phi)}~,\\
    \rho &= \frac 1 2 \dot \phi^2 + V(\phi)~.
    \label{canonical-scalar-w-rho}
\end{align}

We know from cosmological measurements~\cite{Planck:2018vyg} that $w \approx -1$ and $\rho = \Lambda_\mathrm{DE}^4 =  (2.4~\mathrm{meV})^4$~.  Clearly, this requires that we choose our potential such that $V(\phi) \approx \Lambda_\mathrm{DE}^4$, and that the potential is shallow enough such that the field only rolls very slowly $\dot \phi^2 \ll \Lambda_\mathrm{DE}^4$~.

In the case $\dot \phi = 0$, then the theory reduces to a simple cosmological constant.  Any amount of kinetic energy in the field causes the equation of state to deviate from $w = -1$, which then serves as a measure of how dynamical dark energy is.  Through the couplings of Eq.~\eqref{interaction-lagrangian}, it is possible to measure the amount of kinetic energy in the dark energy field via a time-dependent signal in $\mu$.

Previous studies on this topic have focused on a non-minimal coupling to photons~\cite{Barrow:2002zh} (although not in the context of atomic clocks), or suggested using atomic clocks to search for dark energy theories that are coupled to matter via variations in the Higgs vacuum expectation value~\cite{Chakrabarti:2021sgs}.  In the present case, we rely on two ingredients: the non-minimal couplings of Eq.~\eqref{interaction-lagrangian} to generate a scalar field-dependent $\mu$, and a nonzero amount of kinetic energy $\dot \phi$ to cause the field value to change over time.

If we assume that $\dot \phi$ is constant over the timescale of the experiment ($\sim$years) then it is possible to constrain quintessence in a model-independent way.
{Since $\dot \phi$ is constant, over some time period $\Delta t$ the field changes by an amount $\Delta \phi \approx \dot \phi \Delta t$.  Then we have via Eq.~\eqref{mu-variation}
\begin{equation}
    \frac{\Delta \mu}{\bar \mu} = \frac{\dot \phi}{M_\mathrm{eff}}\Delta t \approx \frac{\sqrt{2} \Lambda_\mathrm{DE}^2}{M_\mathrm{eff}} \sqrt{1 + w} \Delta t ~,
\label{delta_mu-DE}
\end{equation}
where in the second equality we have used Eq.~\eqref{canonical-scalar-w-rho} and $V(\phi) \approx \Lambda_\mathrm{DE}^4$.
The maximum allowed deviation from $w = -1$  for a canonical scalar field model of dark energy, as measured by Planck, is $-1 \leq w < -0.95$ at the 95\% confidence level~\cite{Planck:2018vyg}.  It should also be noted that there may be a growing preference for dynamical dark energy from baryon acoustic oscillations~\cite{DESI:2024mwx}.}

\begin{figure}[t]
    \centering
    \includegraphics[width=\linewidth]{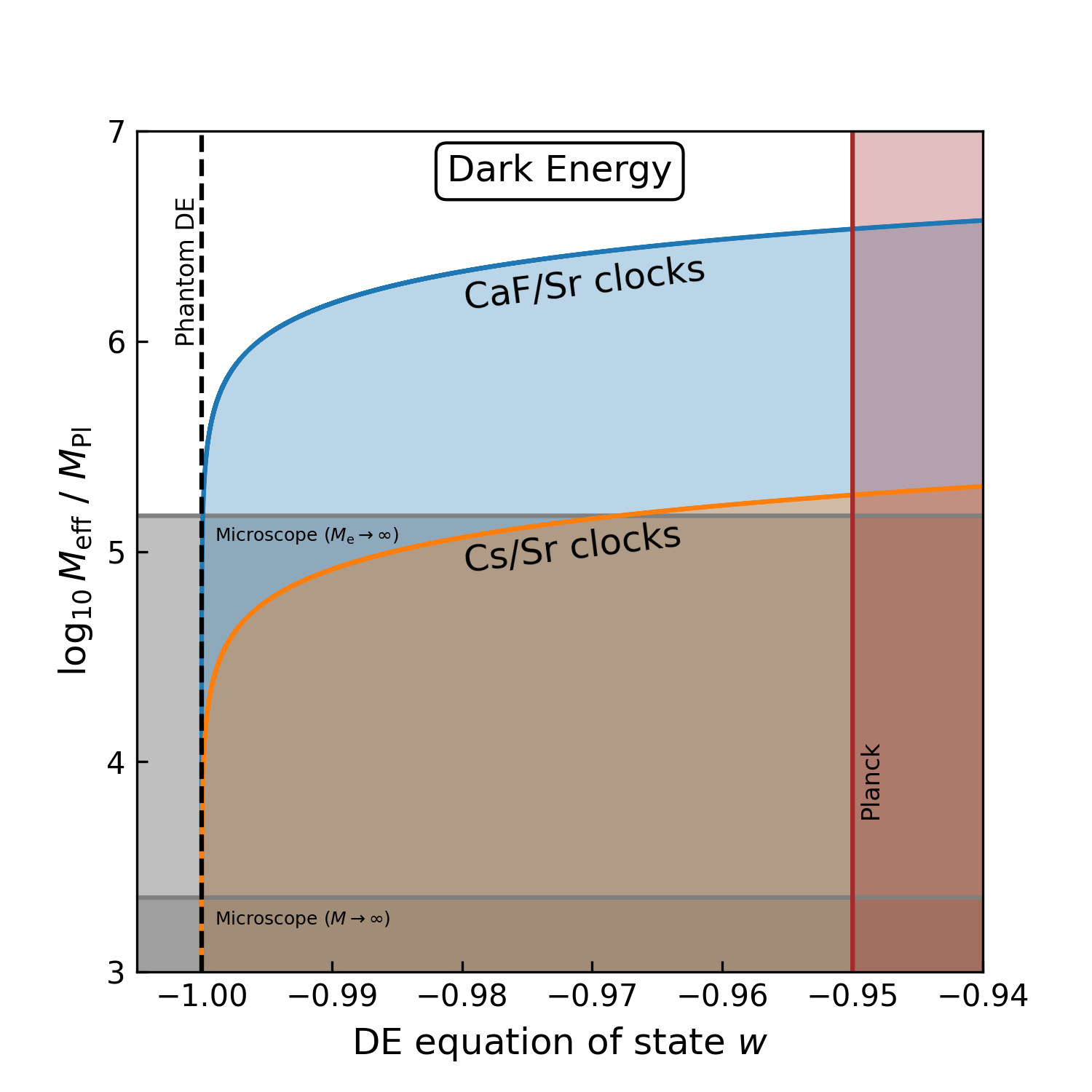}
    \includegraphics[width=\linewidth]{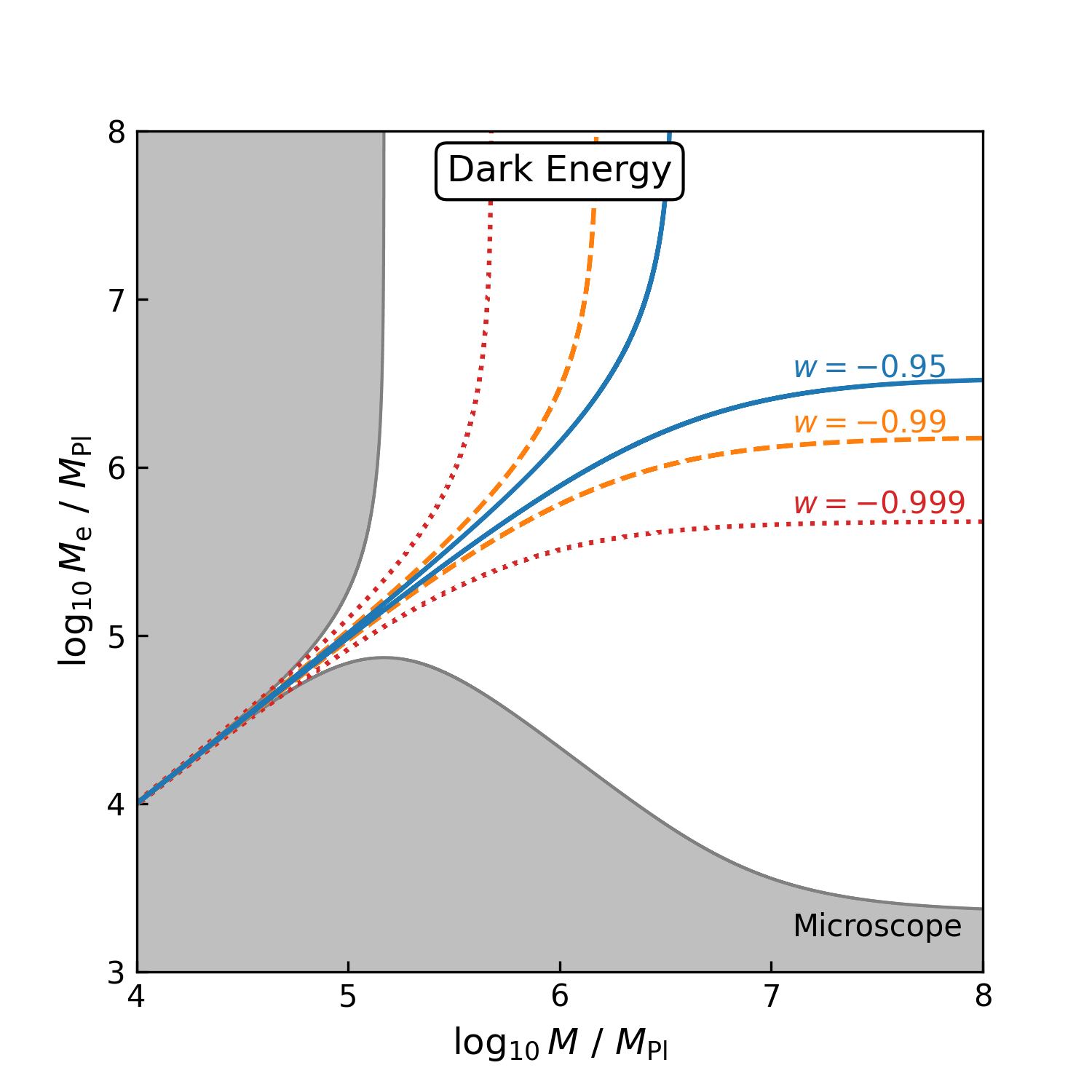}
    \caption{Projected bounds from clocks on quintessence dark energy, over an observation time of 3 years. A CaF/Sr clock pair was used for the bottom plot.}
    \label{fig:dark-energy}
\end{figure}

The scalar field is assumed to be essentially massless on the scale of the Solar System.  As such, tests of the equivalence principle and fifth force tests constrain this theory tightly.  Assuming equivalence principle violation (as we must in order to measure any change in $\mu$ with clocks) then the strongest experimental bound comes from MICROSCOPE, which is derived in Appendix~\ref{sec:microscope}.  A comparison of the expected constraints from atomic clocks to the bounds from MICROSCOPE and Planck are shown in Fig.~\ref{fig:dark-energy}.  The clock bounds strengthen as $w$ increases away from $w = -1$ because this corresponds to more kinetic energy in the dark energy field, giving a larger change to the electron-proton mass ratio over a given period of time.

\subsection{Dark matter}
\label{sec:constraints-dark-matter}

Dark matter may be modeled as a canonical scalar field with a mass $m$:
\begin{equation}
    {\cal L}_\phi = - \frac{1}{2} (\partial \phi)^2 - \frac{1}{2} m^2 \phi^2~.
    \label{DM-lagrangian}
\end{equation}
Cosmologically, we once again assume $\phi = \phi(t)$ such that dark matter is a perfect fluid with an equation of state $w_\mathrm{DM} = 0$ and present-day density $\rho_\mathrm{DM} \approx 0.3 \rho_0$, where $\rho_0$ is the cosmological energy density.
The equation of motion following from Eq.~\eqref{DM-lagrangian} is
\begin{equation}
    (\Box - m^2) \phi = 0~,
    \label{DM-eom}
\end{equation}
where we are neglecting Hubble drag as well as the Standard Model couplings for the time being.  This is a good approximation on short timescales compared to the Hubble time and in regions where the Standard Model field energy densities are small, such as in the vacuum of space.  Cosmological solutions depend only on time, so the solution to Eq.~\eqref{DM-eom} follows as
\begin{equation}
    \phi(t) = \frac{\sqrt{2 \rho}}{m} \cos(m t + \delta)~.
    \label{dm-background}
\end{equation}
The normalization is chosen such that $\rho$ is the energy density of the oscillating background field.  Using Eq.~\eqref{canonical-scalar-w-rho} we find an oscillating equation of state $w = \cos 2 m t$.  Although this equation of state oscillates between $-1$ and $1$, so long as the period $m^{-1}$ is short compared to the Hubble time the oscillations average to zero, giving $w = 0$ on cosmological timescales as desired.  The average value of the dark matter density in the universe is $\rho_\mathrm{DM} \approx 10^{-11}~\mathrm{eV}^4$.  However, this is not the relevant quantity in our Solar System.  The local dark matter density is significantly higher at approximately $\rho_\mathrm{DM, local} \approx 2.6 \times 10^{-6} ~\mathrm{eV}^4$~\cite{Graham:2015ifn}.  In fact, $\phi(t)$ is much more complicated within the galaxy, but over timescales shorter than the coherence time and length scales shorter than the coherence length, it still takes the form of Eq.~\eqref{dm-background}, but with the local dark matter density $\rho_\mathrm{DM, local}$.  The coherence time and coherence length are given by~\cite{Marsh:2022gnf, Sherrill_2023}
\begin{equation}
    \tau_\mathrm{C} = \frac{4 \pi}{m v_\mathrm{vir}^2}~, \quad
    \lambda_\mathrm{C} = \frac{2 \pi}{m v_\mathrm{vir}}~,
    \label{coherence-time}
\end{equation}
where $v_\mathrm{vir} \approx 200~\mathrm{km~s}^{-1}$ is the virial velocity of our galaxy.  The maximum dark matter mass we consider is $m \approx \frac{1}{10~\mathrm{min}} \approx 10^{-18}~\mathrm{eV}$.  This corresponds to a coherence length of $\lambda_\mathrm{C} \approx 10^{12}~\mathrm{km}$, which is larger than the Solar System.  The corresponding coherence time is $\tau_\mathrm{C} \approx 240~\mathrm{years}$, which is much longer than the measurement time of a few years.  As such the whole mass range of dark matter particles under current consideration is well within the regime of applicability of Eq.~\eqref{dm-background}.

\begin{figure}[h!]
    \centering

    \includegraphics[width=\linewidth]{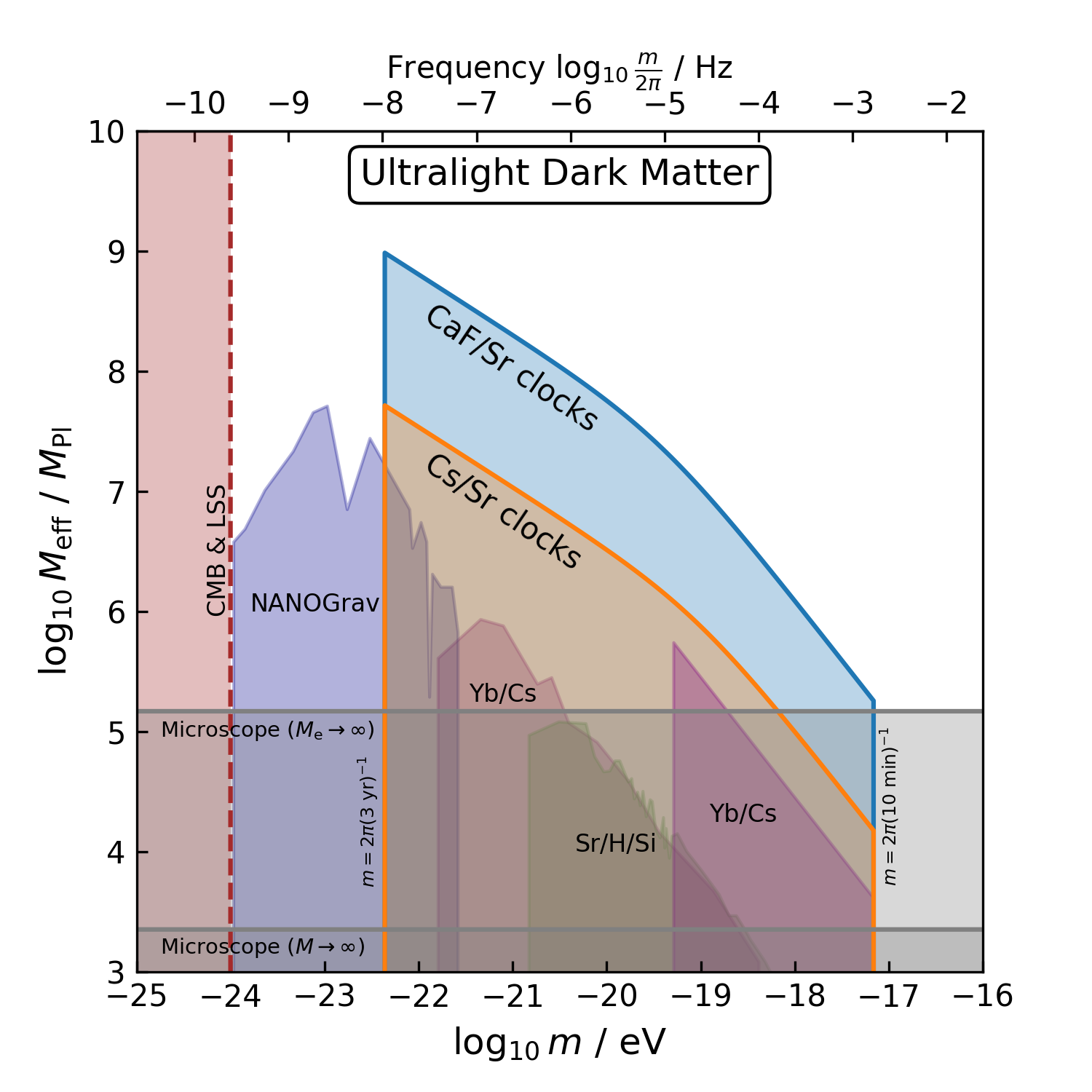}
    \caption{\raggedright Constraints on the space of parameters $(M,m)$ in the Dark Matter model defined in \eqref{dm-background} after marginalizing over the unknown phase $\delta$, for an observation time of $T = 3~\mathrm{yr}$.
    Also plotted are constraints from NANOGrav~\cite{NANOGrav:2023hvm} and Yb/Cs clocks~\cite{Kobayashi_2022, Sherrill_2023} which were drawn from~\cite{AxionLimits}, as well as Sr/H/Si clocks~\cite{Kennedy:2020bac}, and the MICROSCOPE satellite~\cite{MICROSCOPE:2022doy}.  The best torsion balance curves~\cite{Adelberger:2003zx} sit approximately one order of magnitude below the MICROSCOPE line and hence are not included in the figure.  Planned atom interferometry experiments will also be sensitive to the higher end of this mass range ($10^{-19}~\mathrm{eV} \lesssim m \lesssim 10^{-11}~\mathrm{eV}$) within the next few years~\cite{Buchmueller:2023nll}.}
    \label{fig:dark-matter}
\end{figure}

To compute the resulting variations in $\mu$ we combine Eqs.~\eqref{mu-variation} and \eqref{dm-background} and find
\begin{equation}
    \frac{\Delta \mu}{\bar \mu} = \frac{1}{M_\mathrm{eff}} \frac{\sqrt{2 \rho_\mathrm{DM}}}{m} \cos\left(m t + \delta \right)~.
    \label{dm-mu-variation}
\end{equation}
We see that this is of the form $\Delta \mu / \bar \mu \sim \sin(\omega t) / \omega$.  Signals of this form were searched for in the simulated clock data, with the result given in Table~\ref{tab:scalar_variation}.  Applying that result to the fundamental theory parameters $M_\mathrm{eff}$ and $m$ gives the projected constraints in Fig.~\ref{fig:dark-matter}.

Also included in that plot is a comparison to existing constraints, the strongest of which come from CMB and large scale structure, the MICROSCOPE satellite, and existing clock constraints.  Because the signal oscillates with frequency $m$, the maximum and minimum timescales over which data is gathered are directly proportional to the range in $m$ that may be constrained with clocks.

\section{Forecast Tool}
\label{sec:tool}

In this Section we present a software package which can be used to extract constraints for the three signals described in Section \ref{sec:theory_models}, and was used in the creation of this work. The tool takes in a set of clock characteristics, a phenomenological model of new fundamental physics, generates a simulated data stream, and applies the MCMC method as described in Sec.~\ref{sec:simulated_data} to perform parameter estimation.  This forecasts the maximum amplitudes for signals in the time variation of $\mu$ that are associated with new physics, specifically dark energy, dark matter, and modified gravity. The results are then used to generate constraint plots for the different theory scenarios.

The key parameters that must be provided by the user are:
 {\begin{itemize}
    \item The value of the Allan variance for the instability, the accuracy and the sensitivity coefficient $K$ for all the clocks used,
    \item A list of clock pairs to be considered and
    \item The functional form of the phenomenological models one wishes to forecast in the frequency domain.  The models used investigated in this work are included.
\end{itemize}}
The software then generates a set of model-independent constraints for the desired signals in $\mu(t)$.  We have included the analysis to map those variations onto constraints on the theories.

This setup is designed to be extended to a wide variety of EP-violating theories from arbitrary clock characteristics.  {This is demonstrated in the following example.  Let us assume the user wishes to evaluate the capabilities of two new clocks nicknamed ``C1'' and ``C2''.  The user adds a line for each clock in stats/clocks\_parameters.csv, supplying the information on the instability, accuracy and constant K for each clock.  Furthermore, the user specifies the clock combination C1/C2 in stats/clocks\_pairs.csv.  The user then runs the statistical analysis via the supplied makefile, simply running the command ``make'' from the command line.  This generates a new line in the stats/sigma\_A\_table.csv file, giving the uncertainty in the amplitudes for each of the generalized signals in Table~\ref{tab:scalar_variation}.  These can also be queried programatically via an API that is detailed in the code's documentation, and was used in the generation of the theory bound figures in this work.  If desired, one could replace the Cs/Sr clock projections with the C1/C2 clock projection by replacing the clock pair name in each of the plotting scripts supplied with the software package.}
The software package is publicly available\footnote{\url{https://github.com/elizabeth-pa/clock-constraints}} and the version used for this work is permanently archived at \cite{zenodo}.

\section{Conclusions}
\label{sec:conclusions}

Atomic clocks are powerful tools for testing theories involving modified gravity, dark energy, and dark matter by measuring time-dependent changes in fundamental constants like the proton-to-electron mass ratio ($\mu$). Modern atomic clocks have reached an uncertainty of as low as one part in $10^{19}$, allowing for the possibility of probing these fundamental constant variations and discovering new physics beyond the Standard Model. 

In this paper, we highlighted several theories that couple differently to protons and electrons, leading to variations in $\mu$. These couplings cause variations in the atoms' transition frequencies, which clocks are sensitive to.  The signals, coming from fundamental physics, generally manifest as either a linear drift or sinusoidal oscillations in $\mu$.


A preliminary analysis was conducted using publicly available data from the BIPM Circular T database. 
We found that the Circular T constraints were generally sub-leading compared to existing constraints for the theories under investigation.  This work also presented forecasts of the sensitivity of state-of-the-art atomic clocks using the Fisher matrix calculation and an MCMC-based analysis pipeline that utilises simulated data.

This simulated signal was then used to project constraints on a range of fundamental physics theories, highlighting classes of theories where atomic clocks would significantly improve upon existing constraints.  Those improvements are most profound for dynamical dark energy and dark matter theories.  For modified gravity theories, we find an improvement over MICROSCOPE bounds provided that the fifth force falls off more slowly than $1/r^2$.  In this case bounds from lunar laser ranging are currently stronger still, although this situation may change as atomic clock measurements improve.

The analysis framework has been turned into a publicly available tool that can be used by clock operators and researchers to test their clocks' sensitivity to the different theoretical models. The tool enhances the ability to test various models and plan for future experiments. It can also be used to investigate the effects of different clock characteristics, including the level of noise of the clocks, on the ability to detect or further constrain new physics signals. It also presents the possibility to compare the effects of different noise components to allow for informed decision-making in experimental design.  This is important for e.g. detecting these fundamental physics signals by taking specific steps to suppress the noise.

Improvements in atomic clock precision and data coverage and quality could significantly enhance the search for time-varying signals of new physics. For the detection of modified gravity signals where it is hypothesised that the period of oscillation is one year, one would need a dataset spanning at least 2-3 years. Gaps in the data (which are common in current datasets from clocks) also cause a limitation when turning to data analysis tools to infer the presence of a signal. The analysis framework presented is, therefore, a good starting point for testing the capabilities of different atomic clocks and motivating the need for atomic clock readouts that are continuous and as precise and accurate as possible.

In this work, we focused on deriving projected constraints from optical, microwave and molecular clock transition combinations with a focus on the combinations that produce the best constraints in time-dependent variations of $\mu$. As technology progresses and new technological platforms for the development of clocks are explored, new types of clocks have surfaced that have increased sensitivity to time-dependent variations of $\alpha$. Highly charged ion (HCI) clocks such as the californium ion clocks ($Cf^{15+}$ and $Cf^{17+}$) \cite{porsev_2020} have emerged as excellent candidates for searching for variations in $\alpha$ due to their high sensitivities ($K_\alpha = 47$ and $K_\alpha = -43.5$ respectively \cite{clocknetwork}), long excited clock transition lifetimes and their convenient optical transition frequencies \cite{berengut_2010, kozlov_2018}. Another promising candidate for detecting $\alpha$-variations is nuclear clocks \cite{Peik_2021}. Significant progress has been made in the development of a Thorium-229 nuclear clock \cite{Zhang2024}. This nuclear transition has an enhanced sensitivity to temporal variations of $\alpha$ with an enhancement factor for $\alpha$ variation of $-(0.82 \pm 0.25) \times 10^4$ \cite{thorium_sensitivity} and a projected fractional instability of $\sim 10^{-19}$ \cite{Kazakov_2012}. The analysis could be extended to variations of $\alpha$ and include these different species of atomic clocks in the future.

In future work, it would be interesting to analyze the capabilities of atomic clock pairs in a highly elliptical orbit around the Earth or Sun.  Such a configuration would produce much stronger local variations to atomic frequencies from theories of modified gravity, and would also enable the possibility to map the modified gravity force as a function of distance from the Sun.  In the future, it would also be interesting to connect the dark energy bound to specific models of dark energy, along with the bounds of the cosmological variation of particle masses.  Additionally, there is the possibility of searching for multiple phenomenological signals present together in the clock measurement datastream. In this case, one could exploit the different time dependence of the signals to disentangle their contributions and perform a parameter estimation on the full set of parameters for all three signals. For the three phenomenological signals considered in this work, a full reconstruction of all the parameter space is possible\footnote{With one caveat: if the frequency and phase of the dark matter modulation are close to the period and phase of the modified gravity effect due to the Earth's motion around the Sun, the two signals are degenerate and their disentanglement would be virtually impossible.}, as the signals have a very different time dependence (which is even more different in the frequency domain) and their individual effect can therefore be disentangled.
Given that, we expect that the uncertainty in the reconstruction of all the parameters is expected to be slightly worse than the individual constraints of the present work, while constraining all three phenomenological models simultaneously.

{\small {\bf Acknowledgements}:  The authors are grateful to Hannah Banks, Clare Burrage, and Lam Hui for helpful conversations. CdR, BE and AJT are supported in part by the
STFC Consolidated Grants ST/T000791/1 and ST/X000575/1. CdR and BE are also supported in parts by a Simons Investigator award 690508. GM acknowledges support from the Imperial College London Schr\"odinger Scholarship scheme. OB acknowledges support from a Royal Society Leverhulme Trust Senior Research Fellowship.  EP, CFAB, RH, and OB acknowledge support from the AION grant ST/T006994/1, awarded by EPSRC and STFC as part of the Quantum Technology for Fundamental Physics programme. EP, CFAB, and RH received additional support from the USOC grant EP/Y005163/1, awarded by EPSRC.}

\newpage
\appendix

\onecolumngrid

\section{From Circular T data to $\Delta \mu / \mu$}
\label{app:circular-t}

Circular T reports the fractional deviation $d$ of the scale interval of the International Atomic Time (TAI) with respect to the SI second on the geoid (or Terrestrial Time (TT)). This quantity reflects how much the length of one second in TAI differs from the SI second (in the time domain). This quantity relates to the fractional frequency deviation of TAI as follows:
\begin{align}\label{d_def}
d = -y_{\mathrm{TAI}} &= \frac{\text{length of 1s in TAI} - \text{length of 1s in TT}}{\text{length of 1s in TT}}\,.
\end{align}

The values of $d$ for individual Primary Frequency Standards (PFS) and Secondary Frequency Standards (SFS) contributions are also published. These values denote the second as realized by each standard, and they compare TAI frequency with that of the given Primary and Secondary Frequency Standards (PFS/SFS). These are found in Section 3 of the Circular T datasheets~\cite{explanatory_supplement}. We define the value of $d_i$ for individual standards as:
\begin{align}
d_{i} \approx -y_{i} = \frac{f_i}{f_\text{TAI}} - 1 \,,
\end{align}
with
\begin{equation}
f_{i} \coloneqq \frac{\nu_{i}}{\nu_{i,0}},
\end{equation}
where $\nu_{i}$ is the measured frequency and $\nu_{i,0}$ is the nominal frequency of the atomic transition.

We will only use the values for $d$ for optical atomic clocks, i.e. strontium and ytterbium, for which $K_\mu=0$. We assume that the dominant contribution to $f_{\rm TAI}$ is from Caesium-based clocks that use a microwave transition with $K_\mu = -1$. In this way, using Eq.~\eqref{eq:sensitivity} we can assume that $ \frac{\Delta\mu}{\mu} = - d $ (based on the sensitivities of optical and microwave clocks to variations of $\mu$ (Table \ref{tab:clock_stab_acc}). 

Whether a clock contributes to Circular T depends on whether it runs in a given month and whether data is submitted for the calculation of UTC. Secondary Frequency Standards based on optical transitions do not typically contribute as regularly as microwave standards, leading to long gaps sometimes of many months between datapoints. 
A graphical representation of all the Primary and Secondary Frequency Standards evaluations reported since September 2003 is shown on the BIPM website \footnote{\url{https://webtai.bipm.org/database/show_psfs.html}}.

\begin{figure}[h]
  \includegraphics[width=0.8\columnwidth
]{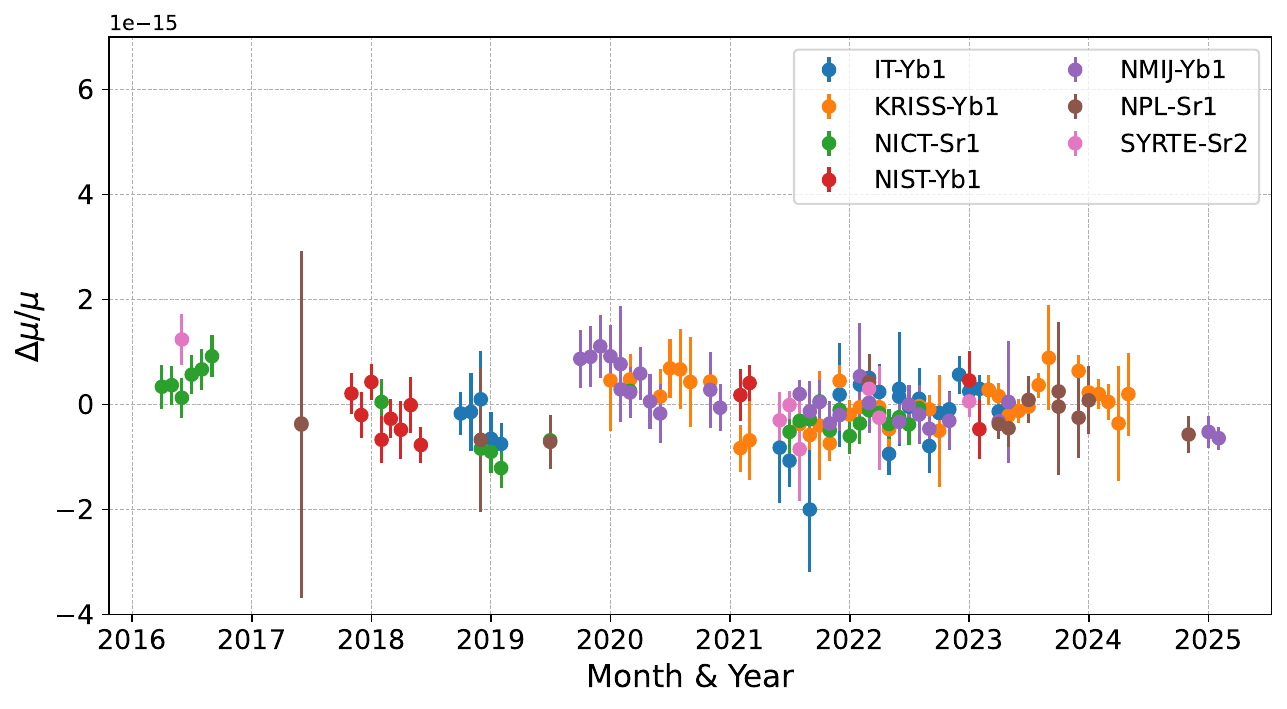}
  \caption{  
  Plot of $\frac{\Delta \mu}{\mu}$ (= $- d$) vs month and year for each optical clock used in the analysis.}
  \label{fig:dvalues}
\end{figure}

One value for $d$ is recorded for each frequency standard, i.e. each atomic clock, per month (if the clock was running that month). Fractional uptimes for each clock are also reported monthly (section 3 of the datasheets). Up-to-date tables containing all the available data for each clock can be found on the BIPM website \footnote{
\url{https://webtai.bipm.org/ftp/pub/tai/other-products/taipsfs/}}. Figure~\ref{fig:dvalues} shows the values for $d$ for all the optical clocks reported in Circular T up to the time of writing. The value of $d$ is reported along with several uncertainty values, which are then summed in quadrature to obtain the final error reported for the values of $d$. The uncertainties reported are that of the instability of the standard, the combined uncertainty from systematic effects, uncertainties in the link between the TAI-participating clock and the standard relating to dead time and systematic effects and the uncertainty in the link to TAI \cite{explanatory_supplement}.

The date in Circular T is reported in Modified Julian Date MJD (start and end for each period of estimation). MJD gives the number of days since midnight on November 17 1858. For the modified gravity case, the phase of the sinusoidal signal is defined by the time that aphelion/perihelion happens during the year. Perihelion and aphelion times vary from year to year. Table~\ref{table1} gives an average time of when perihelion and aphelion occur during the year and the corresponding phase in terms of days since the beginning of the MJD calendar.

\begin{table}[h]
\begin{tabular}{rll}
\toprule
 & Perihelion &   Aphelion \\
\midrule
Date & 4 January & 5 July \\
Phase & 48 days & 230 days \\
\bottomrule
\end{tabular}
\caption{\justifying Average time of the year that perihelion and aphelion happen. The number of days from the beginning of the MJD calendar to the first instance of perihelion and aphelion are also reported as phases.}

\label{table1}
\end{table}

The phase could be set to 48 days but since we have data from 2016 onwards, the MJD day that the perihelion happens in the data (4 January 2017 in MJD = 57757) can be subtracted from the data and then fit the equation without the phase.

\section{Current Clock Characteristics}
\label{app:current_clocks}

Table \ref{tab:clock_stab_acc} presents different species of clocks and their sensitivity to variation in $\alpha$ and $\mu$ along with the clock instability and accuracy (systematic uncertainty).

There are two types of errors in frequency standards: (a) statistical errors originating from measurement fluctuations and (b) systematic uncertainties \cite{clocks_review}. The performance of the clock is characterized by determining the (1) fractional frequency instability $\sigma_y (\tau)$ and (2) systematic frequency uncertainty. The clock stability is limited by the quantum projection noise or shot noise and thermal noise. The systematic frequency fluctuations originate from magnetic and electric fields, which induce perturbations in the natural frequencies of the atoms (e.g. Zeeman and Stark shift). More information on how noise is characterized in clocks can be found in~\cite{clocks_review, nist_stab_handbook}.

The major noise sources can be modeled as white noise and pink (flicker) frequency noise. The total noise of the system can be derived by summing all the noise terms. 

In the frequency domain, the one-sided Power Spectral Density (PSD) function for fractional frequency fluctuations of the clock is:
\begin{equation}
S_y(f) = h_0 + \frac{1}{f} h_{-1}\,.
\label{PSD_equation}
\end{equation}
where $h_0$ is the white frequency modulation (WFM) component (white noise) and the $\frac{1}{f} h_{-1}$ is the flicker frequency modulations (FFM) or pink noise component.

\begin{figure}[]
  \centering
  \begin{subfigure}{0.49\linewidth}
    \centering
    \includegraphics[width=\linewidth]{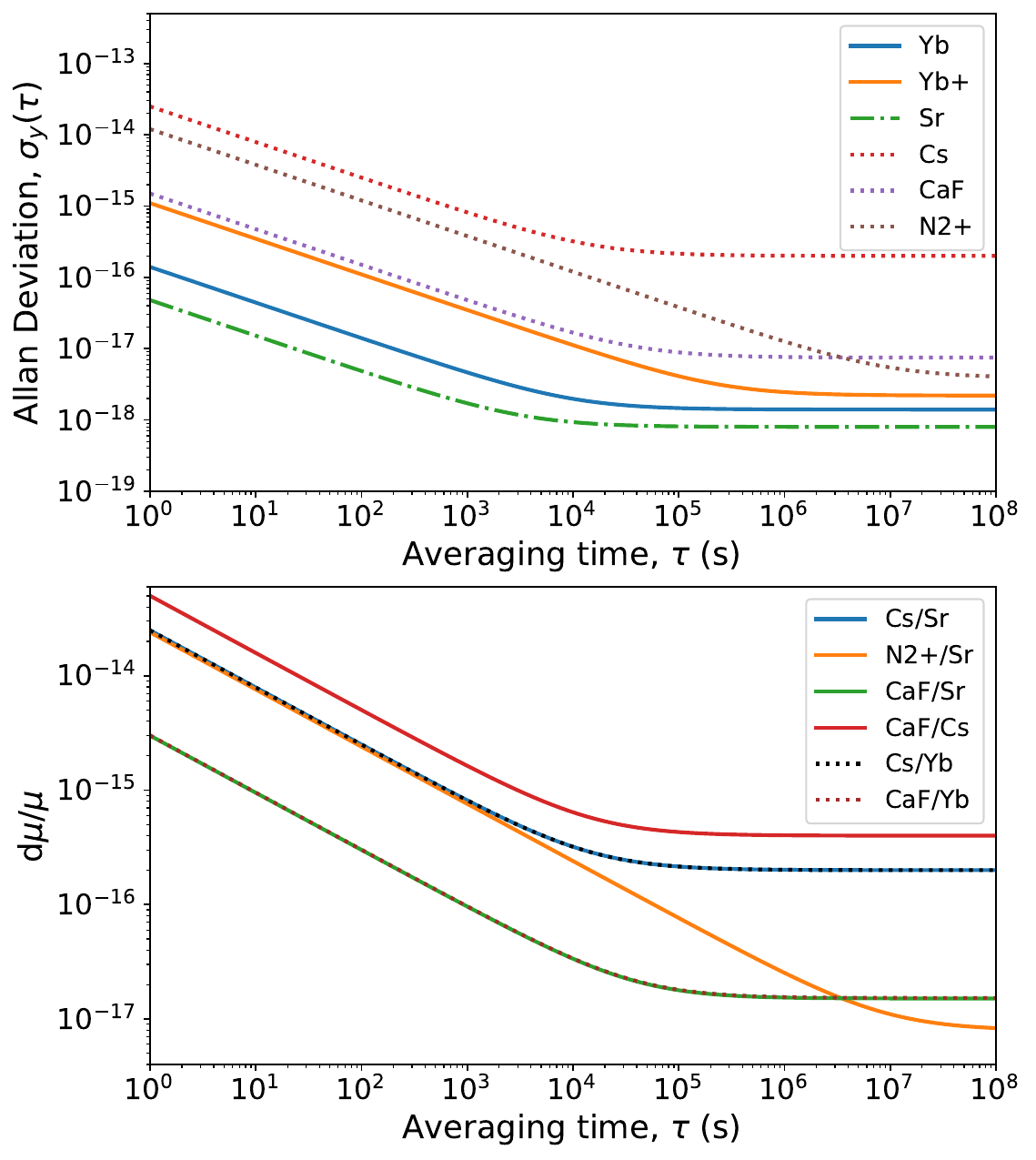}
    \caption{Allan deviation as a function of averaging time.}
    \label{fig:sigma_sensitivity}
  \end{subfigure}
  \hfill
  \begin{subfigure}{0.49\linewidth}
    \centering
    \includegraphics[width=0.905\linewidth]{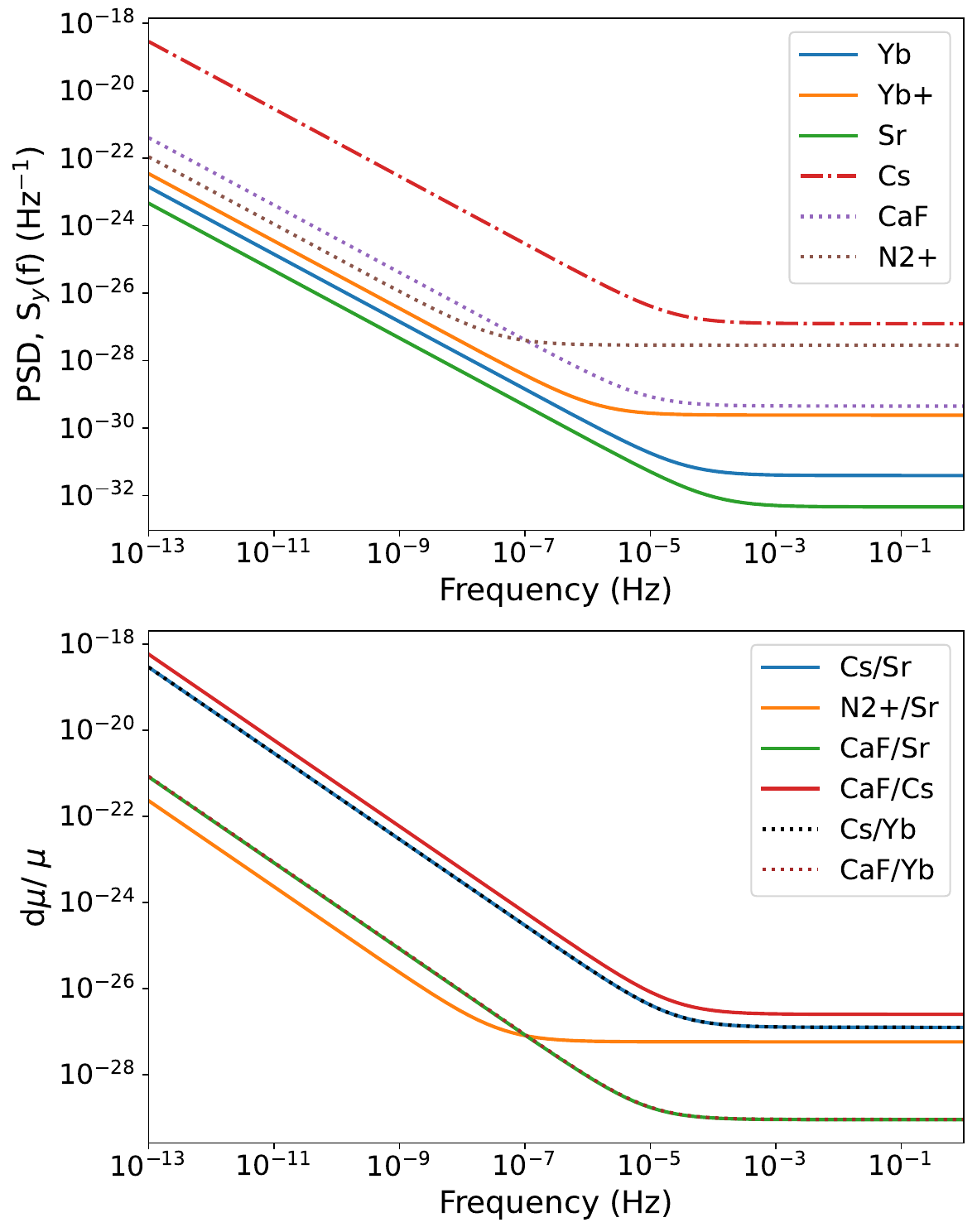}
    \caption{Power Spectral Density as a function of frequency.}
    \label{fig:psd_sensitivity}
  \end{subfigure}
  \caption{(a) Allan deviation (fractional uncertainty) as a function of averaging time for different clock species. The sloped region represents white noise fluctuations (instability), while the flat region corresponds to flicker/pink noise (inaccuracy). The bottom plot illustrates sensitivities to variations in $\mu$ over time for different clock combinations.
(b) Power Spectral Density as a function of frequency for different clock species. The flat region represents white noise fluctuations (instability), while the sloped region corresponds to flicker/pink noise (inaccuracy). The bottom plot shows sensitivities to variations in $\mu$ as a function of frequency for different clock combinations.
Note: In both bottom plots, the Cs/Sr and Cs/Yb curves overlap, as do the CaF/Cs and CaF/Yb curves.}
  \label{fig:combined_sensitivity}
\end{figure}

The relationship between Allan variance and PSD coefficients $h_0$ and $h_{-1}$ depend on the power law model of the noise. The conversions are given by:
\begin{equation}
\begin{aligned}
    \sigma_y^2(\tau) &= \frac{1}{2\tau} h_0\,, \\
    \sigma_y^2(\tau) &= 2 \ln{(2)} h_{-1}\,.
\end{aligned}
\label{PSD_conversion}
\end{equation}
Plots of the combined PSD of the clocks listed in Table \ref{tab:clock_stab_acc} are presented in Fig.~\ref{fig:psd_sensitivity}. The sensitivity of different pairs of clocks to variations in $\mu$ is also presented as a function of frequency. 

\begin{table}[h]
    \centering
    \begin{tabular}{|>{\centering\arraybackslash}m{2cm}|>{\centering\arraybackslash}m{2.5cm}|>{\centering\arraybackslash}m{2.5cm}|>{\centering\arraybackslash}m{2.5cm}|>{\centering\arraybackslash}m{2.5cm}|>{\centering\arraybackslash}m{1.5cm}|>{\centering\arraybackslash}m{1.5cm}|}
        \hline
        \rowcolor{lightgray} Clock & Instability / $\sqrt{\tau / s}$ & $h_0$ / s & Accuracy & $h_{-1}$ & $K_{\alpha}$ & $K_{\mu}$ \\ \hline
        \rowcolor{cyan!20} Yb \cite{Kriss_Yb} & $1.4 \times 10^{-16} $ \cite{schioppo2017} & $4.0 \times 10^{-32}$ & $1.4 \times 10^{-18}$ \cite{McGrew2018} & $1.4 \times 10^{-36}$ & 0.37 \cite{Yb_alpha} & 0 \\ \hline
         \rowcolor{cyan!20} $^{171}\text{Yb}^{+}$ \cite{yb+} & $1.1 \times 10^{-15}$ \cite{PhysRevLett.130.253001} & $2.4 \times 10^{-30}$ & $2.2 \times 10^{-18}$ \cite{Tofful_2024}& $3.5 \times 10^{-36}$ & $-5.95$\cite{Yb_alpha} & 0 \\ \hline
        \rowcolor{cyan!20} $^{87}\text{Sr}$ \cite{Bothwell_2019} & $4.8 \times 10^{-17}$ \cite{Sonderhouse2019} & $4.6 \times 10^{-33}$ & $8.0 \times 10^{-19}$ \cite{best_clock} & $4.6 \times 10^{-37}$ & 0.06 & 0 \\ \hline
        \rowcolor{green!20} $^{133}\text{Cs}$ \cite{cs_npl, cs1} & $2.5 \times 10^{-14}$ & $1.3 \times 10^{-27}$ & $ $1-2$ \times 10^{-16}$ & $2.9 \times 10^{-32}$ & 2.83 & $-1$ \\ \hline
        \rowcolor{purple!20} CaF \cite{clocknetwork} & $1.5 \times 10^{-15}$ & $4.5 \times 10^{-30}$ & $7.5 \times 10^{-18}$ & $4.1 \times 10^{-35}$ & 0 & $-0.5$ \\ \hline
        \rowcolor{purple!20} N$_2^+$ \cite{clocknetwork} & $1.2 \times 10^{-14}$ & $2.9 \times 10^{-28}$ & $3.9 \times 10^{-18}$ & $1.1 \times 10^{-35}$ & 0 & $-0.5$ \\ \hline
    \end{tabular}
    \caption{Clock instability and accuracy of each type of clock used in this analysis. The accuracy is defined as the systematic uncertainty of the clock or flicker noise. The coefficients of the power spectral density of the modeled white and pink noise of the clocks are also stated along with the sensitivity of each clock to variations in $\alpha$ and $\mu$. The instabilities quoted here are based on state-of-the-art clock results except for the CaF and N$_2^+$ clocks where projections of the instability and systematic uncertainty are taken from \cite{clocknetwork}.} 
    \label{tab:clock_stab_acc}
\end{table}

\section{Data analysis and Bayesian inference}
\label{bayesian_inf}

Suppose one has a process which produces some data that are represented by the time series $\{t_i,y_i\}_{i=1...N}$. We also consider that there is an underlying model $h(t,\theta)$, parametrized by the parameters $\theta$, with noise $n$ present as
\begin{align}
y_i=h(t_i,\bar\theta)+n_i\,,
\end{align}
and where we assume Gaussian and uncorrelated noise, i.e. with $\langle n_i \rangle=0$ and $\langle n_i n_j \rangle=\sigma_i^2\delta_{ij}$.
The likelihood probability of the measured $y_i$ as a function of the parameters $\theta$ will be
\begin{align}
\mathcal{L}(\{t_i,y_i\}|\theta)=\prod_i\frac{1}{\sqrt{2\pi\sigma_i^2}}\exp\left(-\frac{(y_i-h(t_i,\theta))^2}{2\sigma_i^2}\right)\,.
\end{align}
If one is then interested in reconstructing the posterior distribution of the parameters $\theta$, namely $P(\theta|\{t_i,y_i\})$, that will be
\begin{align}
P(\theta|\{t_i,y_i\})=\mathcal{L}(\{t_i,y_i\}|\theta)\pi(\theta)\,,
\end{align}
where $\pi(\theta)$ is the prior distribution of the parameters $\theta$.
In principle, one should evaluate the posterior distribution in a space that is highly dimensional. Therefore, Markov Chain Monte Carlo (MCMC) is a very suitable method to produce an efficient representation of this function\footnote{In practice, for a parameter space with dimension $d>3$; it is never convenient to perform a grid evaluation of the posterior.}. The Python package emcee for MCMC \cite{emcee} is the one adopted.

\subsection{Fisher forecast}
\label{sec:fisher}

Given that our instruments can be interrogated each $\sim 1$s, we will have at our disposal a very densely sampled data stream. This allows us to forecast the observability of a variation in time of the frequency ratio $R$ as defined in Eq.~\eqref{eq:sensitivity}. In particular, we promote the datastream to the continuous variable $y(t)$, which comes from the superposition of the modulated signal and the noise
\begin{align}
y(t)=h(t,\theta)+n(t)\,.
\end{align}
For the sake of simplicity, we will assume that the data are observed for a total time $T$ at constant cadency $\Delta t$ and that there are no gaps in the datastream. In this case we also assume a Gaussian, 0-mean, and stationary noise\footnote{The generalization to the case where the first assumption is not taken (i.e. when the data stream is not evenly sampled or it is gapped) involves a time-frequency analysis and is not a complicated one. The treatment of a non-stationary noise can be performed in the time-frequency domain as well.}.
We first Fourier transform the data stream $y(t)$:
\begin{align}
\tilde y(f)=\int_0^T dt\, y(t) \,e^{-2\pi i ft}\,,
\end{align}
and we define the statistics of the noise
\begin{align}
\langle\tilde n^*(f)\tilde n(f')\rangle&=\frac{1}{2}\delta(f-f')N(f)\,,\nonumber\\
N(f)&=h_0+\frac{h_{-1}}{f}\,,
\end{align}
where the matrix $N(f)$ is the noise power spectrum. 
The likelihood function will be
\begin{align}\label{Likelihood_cont}
\mathcal{L}(y|\theta)&=\frac{1}{\mathcal{N}}e^{-\frac 1 2 \chi^2(y,\theta)}\,,\nonumber\\
\chi^2(y,\theta)&=T\int_{\frac{1}{T}}^{\frac{1}{\Delta t}} df\,\frac{\left|\tilde y(f)-\tilde h(f,\theta)\right|^2}{N(f)}\,,
\end{align}
where $\mathcal{N}$ is a normalization factor and $\chi^2(y,\theta)$ the so-called chi-squared. An important note is the fact that when writing Eq.~\eqref{Likelihood_cont} we assumed noise domination (i.e. the amplitude of the signal is less than the noise on average).
In order to produce a forecast, we wish to provide a good summary of the statistics described by the posterior function
\begin{align}
P(\theta|y)=\mathcal{L}(y|\theta)\pi(\theta)\,.
\end{align}
In order to do so, we need to choose a fiducial value for the theoretical parameters $\bar \theta$ and then investigate the structure of the posterior distribution.
Close to its maximum, the chi-squared, peaked at the value of the fiducial parameters $\bar \theta$, can be approximated as
\begin{align}
\chi^2(y,\theta)&\simeq (\theta-\bar\theta)^T\mathcal{F}(\bar\theta)(\theta-\bar\theta)\,,\nonumber\\
\mathcal{F}_{ij}(\bar\theta)&=\frac{1}{2}\frac{\partial^2 \chi^2}{\partial\theta_i\partial\theta_j}(\bar\theta)\,,
\end{align}
where $\mathcal{F}$ is the Fisher Matrix (please note that $\theta$ is a $d$-dimensional vector).
A validation of the Fisher method can be done with a direct comparison of the posterior distribution approximated by the Fisher analysis and the true one by drawing samples from the two distributions and checking that they are compatible. In order to draw samples from the true posterior distribution, an MCMC method is employed as in the released code.

\subsubsection{Fisher forecast for the DM model}

For what concerns the DM signal, which  produces a time modulation of the kind
\begin{align}
\frac{\Delta\mu(t)}{\bar\mu}=\frac{A}{\omega}\sin(\omega t+\delta)
\end{align}
we wish to forecast the detectability of such a signal.
To do so, we first define $\alpha=\frac{A}{\omega}$ for convenience: detection occurs when we are able to reject the hypothesis $\alpha=0$, which is when $\alpha > \sigma_\alpha$, where
\begin{align} \nonumber
\sigma_\alpha &=\mathcal{F}^{-1/2}_{\alpha\alpha}~, \\
&=\left[\int df\,\frac{1}{N(f)}(\text{F}[\cos(\omega t+\phi)](f))^2\right]^{-1/2}~,\nonumber\\
&\simeq\left[\frac{c}{N(\frac{\omega}{2\pi})}\right]^{-1/2}~.
\end{align}
To obtain this result, we used the fact that the discrete Fourier transform of the cosine is well approximated by a delta function with a constant $c$, which is fixed by our units, and $N(f)$ is the noise power spectrum. We are, therefore, able to rule out signals with
\begin{align}
\frac{A}{\omega}&>\sigma_\alpha=\sqrt{\frac{1}{c}N\left(\frac{\omega}{2\pi}\right)}~,\nonumber\\
\Rightarrow A &> \sigma_A\equiv c'\omega\sqrt{h_0+\frac{2\pi\,h_{-1}}{\omega}}~.
\end{align}
The value of the newly defined constant $c'$ can obtained by explicitly computing $\sigma_\alpha=\frac{\sigma_A}{\omega^2}$ at some reference frequency value $\omega_0$.
In Table~\ref{tab:scalar_variation} we compute the values of $\sigma_A(\omega_0)$ for $\omega_0=2\pi\,\text{yr}^{-1}$.
The last formula is valid for values of frequencies $\frac{2\pi}{T_{\rm obs}}\ll\omega\ll \frac{2\pi}{\Delta t}$
with $T_{\rm obs}$ being the total observation time of the experiment and $\Delta t$ the sampling rate.
A further check of the validity of the following formula has been done by an explicit evaluation of $\sigma_A$ at some further values of frequency with MCMC.

\subsection{Realistic gapped signal}
The datastream in a realistic experimental setup will present gaps and a non evenly spaced sampling of the data points, therefore requiring some other steps in the analysis: one cannot simply Fourier transform the observed data stream since gaps will affect the Fourier transform, and the non evenly spacing has to be taken into account in the transform.
In this scenario, the analysis would work as follows
\begin{itemize}
    \item From the observed data stream $y(t)$ one computes $r(t)=y(t)w(t)$ where $w(t)$ is a weight function {(not to be confused with the dark energy equation of state, also referred to as $w$ in Sec.~\ref{sec:projected_constraints})}, which can be set arbitrarily (convenient choices of this weight function will improve the quality of the data analysis but do not change the analysis drastically).
    \item In the generalized frequency domain the datastream will be\\ $\tilde r(f)=\int_{-\infty}^{\infty} df'\, \tilde y(f')\,\tilde w(f-f')$
    \item One has to compute the statistics of the windowed-filtered noise, define the likelihood and produce the forecast exactly as it has been done in this section
\end{itemize}

For the sake of our analysis, given a duty cycle of the experiment $T_{eff}=\beta_{eff}\,T$, with $0<\beta_{eff}<1$, one can replace $T\to T_{eff}$ in the formulas above to obtain a fair proxy of the actual results.

\section{MICROSCOPE}
\label{sec:microscope}
MICROSCOPE is a satellite-borne experiment that measures the differential acceleration between two objects in Earth's orbit,  with one composed of titanium and the other platinum.  The two materials contain slightly different ratios of electrons, protons, and neutrons, so it is sensitive to theories that violate the equivalence principle.  The reported constraint is~\cite{MICROSCOPE:2022doy}
\begin{equation}
    \eta \equiv \frac{2 (a_1 - a_2)}{a_1 + a_2} \lesssim \sqrt{2.3^2 + 1.5^2} \times 10^{-15}\,,
    \label{microscope-constraint}
\end{equation}
where $a_{1, 2}$ are the total accelerations of the titanium and platinum bodies.

The scalar force on a test body is given by Eq.~\eqref{scalar-force}.  It follows that the difference in acceleration between two bodies is, at the lowest order in the electron/proton mass ratio $m_e / m_p$,
\begin{equation}
    a_1 - a_2 = \frac{1}{M_\mathrm{eff}} \left( \epsilon_1 - \epsilon_2 \right) \vec \nabla \phi\,.
\end{equation}
Last, we shall assume that the two objects mainly follow their Newtonian trajectories: $a_1 + a_2 \approx 2 G m_\oplus / R_\mathrm{microscope}^2$, where $R_\mathrm{microscope}$ is the distance between the center of the Earth and the experiment and is $R_\mathrm{microscope} \approx 7000~\mathrm{km}$~\cite{MICROSCOPE:2022doy}.  The parameter $\eta$ is then
\begin{equation}
    \eta = \frac{8 \pi \Mpl^2 R_\mathrm{microscope}^2}{m_\oplus} \frac{1}{M_\mathrm{eff}} \left( \epsilon_1 - \epsilon_2 \right) \phi'\,.
    \label{microscope-bound}
\end{equation}

The external field profile $\phi'$ is sourced by the Earth and is a theory-dependent quantity.
As an example, we specialize to a theory of a massless free scalar, for which Earth's field profile is given by Eq.~\eqref{massless-scalar-external-gradient}.  Then we have, at leading order in $m_e / m_p$,
\begin{equation}
    \eta =  \frac{2 \Mpl^2}{M_\mathrm{eff}} \left( \frac{1}{M} + \frac{\epsilon_\oplus}{M_e} \right) \left( \epsilon_\mathrm{Pt} - \epsilon_\mathrm{Ti} \right)\,.
\end{equation}
The $\epsilon$ factors are given in Table~\ref{tab:epsilon-table}.  In this work, we assume that Earth is mostly composed of iron.

We conclude this Appendix with a few comments.  First, observe that if the scalar couples equally to nucleons and electrons, $M_\mathrm{eff} \to \infty$, then there is no differential acceleration between the test bodies, as expected.  Second, we point out that in the limiting case that the scalar is decoupled from electrons, $M_e \to \infty$, we have $\eta \sim \epsilon_\mathrm{Pt} - \epsilon_\mathrm{Ti}$.  In the opposite limit where the scalar couples to electrons but not nucleons $M \to \infty$, we have $\eta \sim \epsilon_\oplus \left( \epsilon_\mathrm{Pt} - \epsilon_\mathrm{Ti} \right)$~.  This is one higher order in the small ratio $m_e / m_p$, meaning that MICROSCOPE is much less sensitive to scalar fields that couple only to electrons.

\section{Theory model details}
\label{sec:theory}

Our aim in this work is to describe the effects of dark energy, dark matter, and modified gravity on atomic clock measurements.  Although these are disparate phenomena, all can be described by various scalar field theories.  All of the models considered in this work descend from a very general action~\cite{deRham:2010eu}
\begin{equation}
    {\cal L_\phi} = - \frac{1}{2} (\partial \phi)^2 - P(X, \phi) - \frac{1}{\Lambda_3^3} (\partial \phi)^2 \Box \phi - \frac{1}{\Lambda_4^6} (\partial \phi)^2 \left( (\Box \phi)^2 - (\partial_\mu \partial_\nu \phi)^2  \right)\,.
    \label{mother-lagrangian}
\end{equation}
We have omitted the quintic Galileon term, as we will only be concerned with spherically symmetric configurations, for which the quintic Galileon contribution is zero~\cite{Nicolis:2008in}.  We have also included a generic scalar self-interaction potential $P(X, \phi)$, where $X = \frac{1}{2} (\partial \phi)^2$.  If we wished to describe curved spacetimes, the even more general theories of the generalized Galileon~\cite{Deffayet:2009mn} or, equivalently, Horndeski theory~\cite{Horndeski:1974wa} would be the starting point instead or in fact more generically going back to the decoupling limit of the modified gravity theory on that curved background \cite{deRham:2014zqa}.

This theory is coupled to Standard Model particles via Eq.~\eqref{interaction-lagrangian}.  These represent the leading-order low-energy couplings for a real scalar.  If we used a pseudo-scalar instead, such as in the case of an axion (see \cite{Marsh:2015xka} for a review), these would be replaced by couplings to e.g. the fermion pseudoscalar and pseudovector currents and $\tilde F F$.  Couplings to quantities like $\bar e e$ and $F^2$ would only appear at the next order in $\phi/M$ via terms like $\phi^2 F^2$.

In this work, we compute constraints for a single model of dark matter.  Similarly, for dark energy our projected constraints are largely model-independent.  As such, the details of those models are described in Sec.~\ref{sec:projected_constraints}.  For modified gravity, however, there are a wide range of models which must each be translated individually, which is the topic of the majority of this Appendix, after first computing the scalar charge of objects in a model-independent way. Afterwards, we apply the general theory of Eq.~\eqref{mother-lagrangian} to several modified gravity models in turn.  It will often be useful to work with the averaged fermion density fields $\rho_i$.  Assuming that a fermion field $\psi$ is non-relativistic, we have $m_\psi \bar \psi \psi \approx \rho_\psi$, resulting in a more compact notation that will be frequently used throughout this Appendix.

\subsection{Scalar charge}

It is helpful to work on a simple example in detail, which will contain some useful expressions, particularly for the scalar charge of a large source mass.  {We will consider the motion of a test mass $m_\mathrm{test}$ in the presence of a much larger source mass $m_\mathrm{source}$ in the context of a massive free scalar.  The general setup is as follows.  We first compute the field profile sourced by the large source mass, treating the nucleon and electron fields classically such that e.g. $m_p \bar p p = \rho_p$, where $\rho_p$ is the energy density of the protons.  Once the field profile around the source mass is computed, we calculate the motion of the small test mass which is treated as a non-perturbing probe.  (See \cite{Hui:2009kc} for a treatment of such a computation in non-linear scalar field theories.) The Lagrangian for the scalar field is}
\begin{equation}
    {\cal L}_\phi = - \frac{1}{2} (\partial \phi)^2 -  \phi \left( \frac{\rho_p}{M} + \frac{\rho_n}{M} + \frac{\rho_e}{M_e} \right)~,
\end{equation}
where we have converted the couplings of Eq.~\eqref{interaction-lagrangian} to classical energy densities, and we have coupled protons and neutrons with equal strength $M$.  We begin by solving for the external field around the spherical source mass $m_\mathrm{source}$, ignoring the presence of the test mass.  The static equation of motion is
\begin{equation}
    \vec \nabla^2 \phi = \frac{\rho_p}{M} + \frac{\rho_n}{M} + \frac{\rho_e}{M_e}~,
\end{equation}
where the $\rho_i$ are the energy densities of their respective fields. 

Integrating once, we find that the field outside the source is
\begin{equation}
    \phi'(r > R) = \frac{1}{4 \pi r^2} \left( \frac{N_p m_p}{M} + \frac{N_n m_n}{M} + \frac{N_e m_e}{M_e}  \right) \equiv \frac{Q}{4 \pi r^2}~.
    \label{massless-scalar-external-gradient}
\end{equation}
where $N_{p, n, e}$ denotes the total number of protons, neutrons, and electrons in the source mass.  In the second equality we have identified the scalar charge of the source mass:
\begin{equation}
    Q_\mathrm{source} = \frac{N_p m_p}{M} + \frac{N_n m_n}{M} + \frac{N_e m_e}{M_e}~.
\end{equation}
It will be helpful to simplify this expression.  To do so, we assume that the source mass is composed of a single atomic species of atomic number $Z$ and mass number $A$.  The total mass of an individual atom in the source is
\begin{align} \nonumber
    m_\mathrm{atom} &\approx A m_p + Z m_e~, \\
    &= A m_p \left(1 + \frac{Z m_e}{A m_p} \right)~.
\end{align}
where we have ignored the difference in mass between the protons and neutrons.
The combination in the second line will frequently appear, so we define
\begin{equation}
    \epsilon \equiv \frac{Z m_e}{A m_p}~.
    \label{epsilon-ratio}
\end{equation}
This dimensionless ratio quantifies the fraction of the source's mass that is contributed by electrons, and is tabulated for several different atomic species in Table~\ref{tab:epsilon-table}.  We can now rewrite the scalar charge as
\begin{equation}
    Q_\mathrm{source} = \frac{m_\mathrm{source}}{1 + \epsilon_\mathrm{source}} \left( \frac{1}{M} + \frac{\epsilon_\mathrm{source}}{M_e} \right)~.
    \label{scalar-charge}
\end{equation}
Note that the scalar charge defined in this way is a dimensionless quantity.  If we were to couple to nucleons and electrons equally, $M = M_e$, we would have $Q_\mathrm{source} = m_\mathrm{source} / M$, and we recover the familiar expression $\phi'(r) = m_\mathrm{source} / (4 \pi M r^2)$.




\begin{table}[]
    \centering
    \begin{tabular}{|c||c|c|c|c|}
    \hline
       \quad $\epsilon \equiv \frac{Z m_e}{A m_p}^{\phantom Z}_{\phantom P}$ ~ & ~Hydrogen~ & ~~~~Iron~~~~ & ~Titanium~ & ~Platinum~ \\ \hline\hline
        ~$\epsilon$~  & $5.00 \times 10^{-4}$ & $2.33  \times 10^{-4}$ & $2.30  \times 10^{-4}$ & $2.00  \times 10^{-4}$ \\
        \hline
    \end{tabular}
    \caption{\small The dimensionless ratio $\epsilon$, given by Eq.~\eqref{epsilon-ratio}, for the various atomic species that are considered in this work.}
    \label{tab:epsilon-table}
\end{table}

Having solved for the field profile $\phi'(r)$ around our large source mass, we pause briefly to comment on atomic clocks.  We can integrate Eq.~\eqref{massless-scalar-external-gradient} once more to obtain $\phi(r)$, which are treated as probes living in this field profile. Atomic clock tests are sensitive to the ambient field value via the proton/electron mass ratio as in Eq.~\eqref{mu-eff}.  In the case explored in this work, the source mass is the Sun, and clocks on Earth move closer and further away over the course of Earth's orbit.

We now return to the main task of computing the scalar-mediated force between a large source mass and a small test mass.  In the presence of a spatially-varying scalar field profile $\phi = \phi(\vec x)$, a test particle of mass $m_\mathrm{test}$ experiences a scalar-mediated force (see e.g. \cite{Hui:2009kc})
\begin{equation}
    m_\mathrm{test} \vec a_\phi = - \left( \frac{N_p m_p}{M_p} +  \frac{N_n m_n}{M_n} + \frac{N_e m_e}{M_e} \right) \vec \nabla \phi \equiv - Q_\mathrm{test} \vec \nabla \phi~,
    \label{scalar-force}
\end{equation}
where the $N_i$ now refer to the number of constituent particles in the test mass.  In the second equality we have used the same steps as before to define the scalar charge of the test particle, given by Eq.~\eqref{scalar-charge} except with the substitution $m_\mathrm{source} \to m_\mathrm{test}$ etc.
Combining with Eq.~\eqref{massless-scalar-external-gradient}, it follows that the scalar force between our two objects is
\begin{equation}
    \vec F_\phi = - \frac{Q_\mathrm{source} Q_\mathrm{test}}{4 \pi r^2} \hat r~.
\end{equation}
In the course of this work, the source mass will frequently be the Earth, and the test mass will either be test masses in a satellite (for the MICROSCOPE experiment, Appendix~\ref{sec:microscope}) or the moon (for lunar laser ranging, discussed in Appendix~\ref{app:Galileon}).

\subsection{Modified gravity}
\label{app:modified_gravity}

Since the scalar field couples to matter, all objects source a scalar field profile as described in Section~\ref{sec:theory_models}.  Any objects that move in relation to the atomic clocks then cause the ambient field value to change, thus inducing a variation in $\mu$ that can be measured by the clocks.  For long-ranged field profiles, which we are mainly concerned with here, the strongest effect generally comes from the motion of the Sun.

As the Earth orbits the Sun, the Earth-Sun distance varies between $R$ and $R + \Delta R$, where $R \approx 1~\mathrm{au}$ and $\Delta R \approx 0.033 R$\footnote{For 2024, the Earth's closest and furthest distances from the Sun are $0.983~\mathrm{au}$ and $1.017~\mathrm{au}$, respectively.}.
Using Eq.~\eqref{mu-variation}, we see that $\mu$ varies by an amount
\begin{align} \nonumber
    \frac{\Delta \mu}{\bar \mu} &= \frac{\phi(R + \Delta R) - \phi(R)}{M} \,, \\
    &\approx \frac{\phi'(R) \Delta R}{M}~,
    \label{clocks-constraint}
\end{align}
This modified gravity effect manifests as a variation of $\mu$ and $\alpha$ on a one-year period.
To proceed further, we must supply a specific value for $\phi'$, which is model-dependent.  In the remainder of this section, we will consider several different models.

All of the models discussed in this Section are more typically proposed with a universal coupling to matter: $M = M_e$.  However, from a phenomenological point of view, this need not be the case, and we will always consider the case $M \neq M_e$ in order to induce variations in $\mu$.  To keep the discussion as concrete as possible, throughout this Section we shall specialize to the case in which the scalar is decoupled from electrons: $M_e \to \infty$.  It will frequently be necessary to compute the scalar charge of an object, usually the Sun, in a given theory. This is given by Eq.~\eqref{scalar-charge}, and with $M_e \to \infty$, the scalar charge is very nearly determined by its total mass:
\begin{equation}
    Q = \frac{m_\mathrm{obj}}{M} \left(1 + \epsilon \right)^{-1} \approx \frac{m_\mathrm{obj}}{M}~.
\end{equation}
Including the scalar-electron coupling $M_e$ would not change the main results in any meaningful way, nor the size of the projected effects relative to other experimental bounds such as MICROSCOPE.

\subsubsection{Massless free scalar}
The simplest modification of gravity is with a massless free scalar field:
\begin{equation}
    {\cal L} = - \frac{1}{2} (\partial \phi)^2 - \frac{\phi}{M} \rho_\mathrm{nucleon}~,
\end{equation}
where we have included the interaction term with nucleons in the interest of clarity.  (Recall that the energy density of the nucleons, assuming they are non-relativistic, is $\rho_\mathrm{nucleons} \approx m_n \bar n n + m_p \bar p p$.)
Its equation of motion is
\begin{equation}
    \Box \phi = \frac{1}{M} \rho_\mathrm{nucleons} \approx \frac{1}{M}\rho_\mathrm{matter}~.
\end{equation}
We will be interested in the static, spherically symmetric solution around an object of total mass $m_\mathrm{obj}$:
\begin{equation}
    \vec \nabla^2 \phi = \frac{\rho_\mathrm{matter}}{M}
\end{equation}
which may be integrated once to obtain
\begin{equation}
    \phi' = \frac{m_\mathrm{obj}}{4 \pi M r^2}~.
    \label{free-scalar-profile}
\end{equation}
If we use $m_\mathrm{obj} = m_\odot$ then this gives the field profile sourced by the Sun, and $r$ is the distance between the Earth and the Sun.  Putting this into Eq.~\eqref{clocks-constraint}, we rule out
\begin{equation}
    \frac{\Delta \mu / \bar \mu}{10^{-16}} < 6.8 \times 10^6 \left( \frac{\Mpl}{M} \right)^2~.
    \label{massless-free-scalar-bound}
\end{equation}
where $\Mpl = (8 \pi G)^{-1/2} = 2.4 \times 10^{27}~\mathrm{eV}$ is the reduced Planck mass and $\Delta \mu / \bar \mu$ is the maximum variation in from the clock measurement over 1 year.  For example, assuming a clock constraint of $\Delta \mu / \bar \mu < 10^{-16}$ we rule out $M < 10^{3} \Mpl$.

For the MICROSCOPE constraint, we use $m_\mathrm{obj} = m_\oplus$, and $r$ will refer to the distance between the center of the Earth and the MICROSCOPE satellite.  We then substitute the field profile Eq.~\eqref{free-scalar-profile} into Eq.~\eqref{microscope-bound}, finding a constraint
\begin{equation}
    1 < 2.2 \times 10^{10} \left( \frac{\Mpl}{M} \right)^2~.
\end{equation}
In order for clocks to be competitive with MICROSCOPE in this theory, their measured uncertainty of $\Delta \mu / \bar \mu$ would need to be reduced by 3-4 orders of magnitude.

\subsubsection{Massive scalar field}
A slight variation on the previous model involves the addition of a mass for the scalar field:
\begin{equation}
    {\cal L} = - \frac{1}{2} (\partial \phi)^2 - \frac{1}{2} m_\phi^2 \phi^2 - \frac{\phi}{M} \rho_\mathrm{nucleon}~.
\end{equation}
The equation of motion for a static, spherically symmetric field configuration is
\begin{equation}
    \vec \nabla^2 \phi - m_\phi^2 \phi = \frac{1}{M} \rho_\mathrm{nucleon}~,
\end{equation}
which may be integrated once to give
\begin{equation}
    \phi' = \frac{m_\mathrm{obj}}{4 \pi M r^2} e^{- m_\phi r}~.
\end{equation}
Over distances shorter than $r \ll m_\phi^{-1}$, this reduces to the massless field profile Eq.~\eqref{free-scalar-profile}.  So for $m_\phi^{-1} > 1~\mathrm{au}$ the constraint is nearly identical to Eq.~\eqref{massless-free-scalar-bound}.  When $10^4~\mathrm{km} < m_\phi^{-1} < 1~\mathrm{au}$ the clock effect is exponentially suppressed by the $e^{- m_\phi r}$ term, but the MICROSCOPE constraint is still effectively the same as in the massless case.  When $m_\phi^{-1} \lesssim 10^4~\mathrm{km}$, the constraint on $M$ from both experiments is exponentially weaker.

\subsubsection{Galileon}
\label{app:Galileon}

The Galileon around a spherical source of mass $m_\mathrm{obj}$, obeys an equation of motion (after being integrated once)
\begin{equation}\label{eq:Galileon_newparam}
    \left( \frac{\phi'}{r} \right) + \frac{2}{\Lambda^3}\left( \frac{\phi'}{r} \right)^2 + \frac{2 c_4}{\Lambda^6}\left( \frac{\phi'}{r} \right)^3 = \frac{m_\mathrm{obj}}{4 \pi M r^3}~.
\end{equation}
These terms correspond, in order on the left-hand side of this equation, to an ordinary kinetic term, the cubic Galileon terms, and the quartic Galileon terms, respectively.  A quintic Galileon term exists in general, but it vanishes around spherical objects, so testing it would require a different methodology than what is considered in this work.  For perturbations to be stable, it is required that~\cite{Nicolis:2008in}
\begin{equation}
    0 \leq c_4 \leq 2/3~.
    \label{Galileon-stability}
\end{equation}
Around any given object, the solution for $\phi'$ exhibits three distinct regimes, delineated by two Vainshtein radii $R_3$ and $R_4$.  These radii are defined via a function $g(r)$ that quantifies the deviation from the solution for a massless field:
\begin{equation}
    \phi' = \frac{m_\mathrm{obj}}{4 \pi M r^2} g(r)~.
\end{equation}
In terms of $g(r)$, the equation of motion is
\begin{equation}
    g + \left( \frac{R_3}{r} \right)^3 g^2
    + \left( \frac{R_4}{r} \right    )^6 g^3 = 1~.
\end{equation}
This defines the Vainshtein radii as
\begin{align} \nonumber
    R_3 &= \frac{1}{\Lambda} \left( \frac{m_\mathrm{obj}}{2 \pi M} \right)^{1/3}~, \\ 
    R_4 &= \left( \frac{c_4}{2} \right)^{1/6} R_3~.
    \label{vainshtein-radii}
\end{align}
Notice that the stability criterion Eq.~\eqref{Galileon-stability} translates to $R_4 \leq \frac{1}{3^{1/6}} R_3$ for perturbations to be stable.  We are now able to identify three distinct regimes, depending on the distance to the source:
\begin{equation}
    \phi'(r) \approx \begin{cases}
        \Lambda^2 \left(\frac{m_\mathrm{obj}}{8 \pi M c_4} \right)^{1/3} & r < R_4^4 / R_3^3 ~, \\
        \sqrt{ \frac{m_\mathrm{obj} \Lambda^3}{8 \pi M r} } & R_4^4 / R_3^3 < r < R_3~, \\
        \frac{m_\mathrm{obj}}{4 \pi M r^2} & R_3 < r~.
    \end{cases}
    \label{eq:app_Galileon_regimes}
\end{equation}
In other words, very far from the source, the canonical kinetic term dominates.  Moving closer to the source, the cubic Galileon term dominates, and closer still, the quartic Galileon term dominates.  It is worth noting that very close to the source, perturbations in the Galileon field become unstable~\cite{Nicolis:2008in}.  Constraints on this model are discussed in Sec.~\ref{sec:projected_constraints}.

There exist bounds from lunar laser ranging (LLR) on the Galileon~\cite{Nordtvedt:2003pj, Dvali:2002vf, deRham:2016nuf, Tsujikawa:2019pih}, which we briefly summarize following the discussion in~\cite{Babichev:2013usa}.  We begin by noting that the Galileon contributes an additional potential $\delta \Psi = \phi(r) / M$, where $\phi$ is obtained by integrating Eq.~\eqref{eq:app_Galileon_regimes} once.  This potential results in an additional anomalous perihelion precession $\delta \phi_p$ of the moon around the Earth
\begin{equation}
    \delta \phi_p = \pi r \partial_r \left( r^2 \partial_r \left( \frac{\epsilon}{r} \right) \right)~,
\end{equation}
where $\epsilon \equiv \delta \Psi / \Psi_N$ is the ratio of the additional potential to the Newtonian gravitational potential $\Psi_N = G m_\mathrm{obj} / r$.  Substituting, we find
\begin{equation}
    \delta \phi_p = \begin{cases}
    2 \pi \Mpl^2 \Lambda^2 \left( \frac{(8 \pi)^2}{c_4 M^4 m_\mathrm{obj}^2} \right)^{1/3} r^2 & r < R_4^4 / R_3^3 ~, \\
    \frac{3 \pi}{2} \Mpl^2 \left( \frac{8 \pi \Lambda^3 r^3}{m_\mathrm{obj} M^3} \right)^{1/2} & R_4^4 / R_3^3 < r < R_3~, \\
    0 & R_3 < r~.
        \end{cases}
\end{equation}
The experimental bound is $\delta \phi_p \leq 2.4 \times 10^{-11}$.  For the Earth-moon system, we use $m_\mathrm{obj} = M_\mathrm{Earth}$ and set $r$ to the average Earth-moon distance, $r \approx 3.8 \times 10^{10}~\mathrm{cm}$.  The resulting bounds are plotted in Fig.~\ref{fig:MG-Galileon}.
Note that this expression does not account for any effects from the Sun, which could be significant, as will be discussed briefly in the next section.

\subsubsection{Dressing of the Galileon coupling constants}
\label{sec:Galileon-dressing}

The Galileon, along with other theories that exhibit screening, has a strongly nonlinear equation of motion.  As such, it does not obey a superposition principle, so it is not always appropriate to consider a system independently of its environment.  For instance,  when examining the interaction between the Earth and the MICROSCOPE satellite, it may be necessary to consider the effects of the Sun, depending on the theory parameters in question.

Let us briefly re-examine the MICROSCOPE constraint for the cubic Galileon in this context.  We will assume that, in the vicinity of the Earth, we can decompose the field as
\begin{equation}
    \phi(R + r) = \phi_\mathrm{Sun}(R) + \varphi(r)~,
    \label{dressing-field-decomposition}
\end{equation}
where $R$ is the Earth-Sun distance, $r \ll R$ is the distance to the Earth, and $\phi_\mathrm{Sun}$ is the field sourced by the Sun.  This field solution is characterized by the ratio $r_V / R$, where $r_V = R_3$ is the (cubic) Vainshtein radius of the sun, given by Eq.~\eqref{vainshtein-radii}.

The task is to solve for the Lagrangian for the fluctuations $\varphi$ around this background field.  It turns out that the Lagrangian is essentially the same as the original cubic Galileon, but with ``dressed'' (i.e. rescaled) coupling parameters~\cite{Luty:2003vm,Berezhiani:2013dca}:
\begin{equation} \nonumber
    M_\mathrm{eff} = \left(\frac{r_V}{r}\right)^{3/4} M~, \quad \quad
    \Lambda_\mathrm{eff} = \left(\frac{r_V}{r}\right)^{3/4} \Lambda~.
\end{equation}
In the analysis of Sec.~\ref{sec:constraints-modified-gravity}, we included regimes where $r_V / r \gg 1$ so both couplings could, in principle, be dressed to a larger value.  This weakens the matter coupling, but it also weakens the screening effects of the theory.  We can see from Eq.~\eqref{eq:app_Galileon_regimes} that the net effect is to strengthen $\phi'$ by a factor of $r_V / r$.

A proper treatment of this effect would take into account the full cubic Galileon theory, as well as a careful consideration of the Vainshtein and strong-coupling radii, and in principle could result in stronger constraints from MICROSCOPE and lunar laser ranging~\cite{Nordtvedt:2003pj, Dvali:2002vf, deRham:2016nuf, Tsujikawa:2019pih} on this theory.  This detailed analysis is beyond the present scope and will appear in future work.


\subsubsection{DBI--Galileon}
The Galileon enjoys a special class of Galilean symmetries but these be relaxed to a DBI-Galileon classes of interactions ubiquitous from extra dimensions \cite{deRham:2010eu}. The leading term in that framework is given by the simple DBI relativistic kinetic term
\begin{equation}
    {\cal L} = \Lambda^4 \sqrt{1 - \Lambda^{-4}(\partial \phi)^2} - \frac{\phi}{M}\rho_\mathrm{nucleons}~,
\end{equation}
where we have slightly redefined the theory to only couple to nucleons, as the original theory had a universal matter coupling.
The static, spherically symmetric field profile around an object of mass $m_\mathrm{obj}$ is \cite{Goon:2010xh,Burrage:2014uwa}
\begin{equation}
    \phi' = \frac{\Lambda^2}{1 + (r/r_*)^4}
    \quad \quad r_* \equiv \frac{1}{\Lambda} \sqrt{ \frac{m_\mathrm{obj}}{4 \pi M}}~,
\end{equation}
which is written in terms of the Vainshtein radius $r_*$.
Deep within the Vainshtein radius, the solution is approximately
\begin{equation}
    \phi' \approx \Lambda^2~.
\end{equation}
Assuming the Earth is within the Sun's Vainshtein radius ($r_* > 1~\mathrm{au}$) we have
\begin{equation}
    \frac{\Delta \mu / \bar \mu}{10^{-16}} < 1.0 \times 10^5 \left(\frac{\Mpl}{M} \right) \left( \frac{\Lambda}{\mathrm{eV}} \right)^2~.
\end{equation}
Likewise, from MICROSCOPE we have
\begin{equation}
    1 < 2.5 \times 10^5 \left(\frac{\Mpl}{M} \right) \left( \frac{\Lambda}{\mathrm{eV}} \right)^2~.
\end{equation}
We see that the MICROSCOPE bound is comparable to the atomic clock projection, assuming that clocks constrain at the level of $\Delta \mu / \bar \mu \approx 10^{-16}$.  Like the Galileon, it should be subject to an analogous dressing effect of the coupling constants, which will be investigated in future work.

\subsubsection{Generalized interaction}
\label{app:MG_gen_int}

All of the previous theories may be captured with a parametrized description.  In all cases considered so far the field profile may be written in the form
\begin{equation}
    \phi' = \Lambda^2 \left( \frac{m_\mathrm{obj}}{8 \pi M} \right)^\alpha \left( \Lambda r \right)^{-\beta}~.
    \label{generalized-profile}
\end{equation}
What is not encapsulated by this formula is that a single theory may have different scalings for $\phi'$ in different regimes.  For instance, the full DBI-Galileon solution may go as either $\phi' \sim \Lambda^2$ or $\phi' \sim \Lambda^2 (r / r_*)^{-4}$ depending on the size of $r / r_*$, whereas this prescription assumes a single power law scaling with $r$ \cite{Goon:2010xh,Burrage:2014uwa}.  Likewise, this generalized model does not capture the dressing effects mentioned in Sec.~\ref{sec:Galileon-dressing} (see Ref.~\cite{deRham:2014wfa}).  Nevertheless, it provides a useful guide for the general types of interactions that may be tested.

For the MICROSCOPE and LLR constraints, we assume that there is a matter-scalar coupling of the form ${\cal L}_\mathrm{int} = \phi \rho / M$.  It then follows that the acceleration of a test particle is $a = - \vec \nabla \phi / M$.  Following the arguments given in \ref{app:Galileon}, the LLR bound then becomes
\begin{equation}
    \delta \phi_p > \pi (2 - \beta) \left(\frac{\Mpl}{M}\right)^2 \left( \frac{m_\mathrm{obj}}{8 \pi M} \right)^{\alpha - 1} (\Lambda r)^{2-\beta}~.
\end{equation}
We see that this bound vanishes for an inverse square law interaction, $\beta = 2$.
All constraints on this parametrized model are shown in Fig.~\ref{fig:MG_gen_int}.

\subsubsection{Chameleon}
Theories in the chameleon family~\cite{Khoury:2003aq} have a screening mechanism such that an object's scalar charge is not simply a sum of the scalar charges of its constituent particles.  For reviews of the chameleon and related theories, see~\cite{Joyce:2014kja, Burrage:2016bwy, CANTATA:2021asi,Brax:2021wcv}.  In these theories, the scalar charge is suppressed for sufficiently large and dense objects in a way that also depends on the ambient chameleon field value.  The simplest chameleon Lagrangian is
\begin{equation}
    {\cal L} = - \frac{1}{2} (\partial \phi)^2 - \frac{\Lambda^5}{\phi} - \frac{\phi}{M}\rho_\mathrm{nucleon}~.
\end{equation}
The interaction terms are frequently considered together as an ``effective potential'':
\begin{equation}
    V_\mathrm{eff}(\phi) = \frac{\Lambda^5}{\phi} + \frac{\phi}{M} \rho~.
\end{equation}

\begin{figure}
    \centering
    \includegraphics[width=0.45\linewidth]{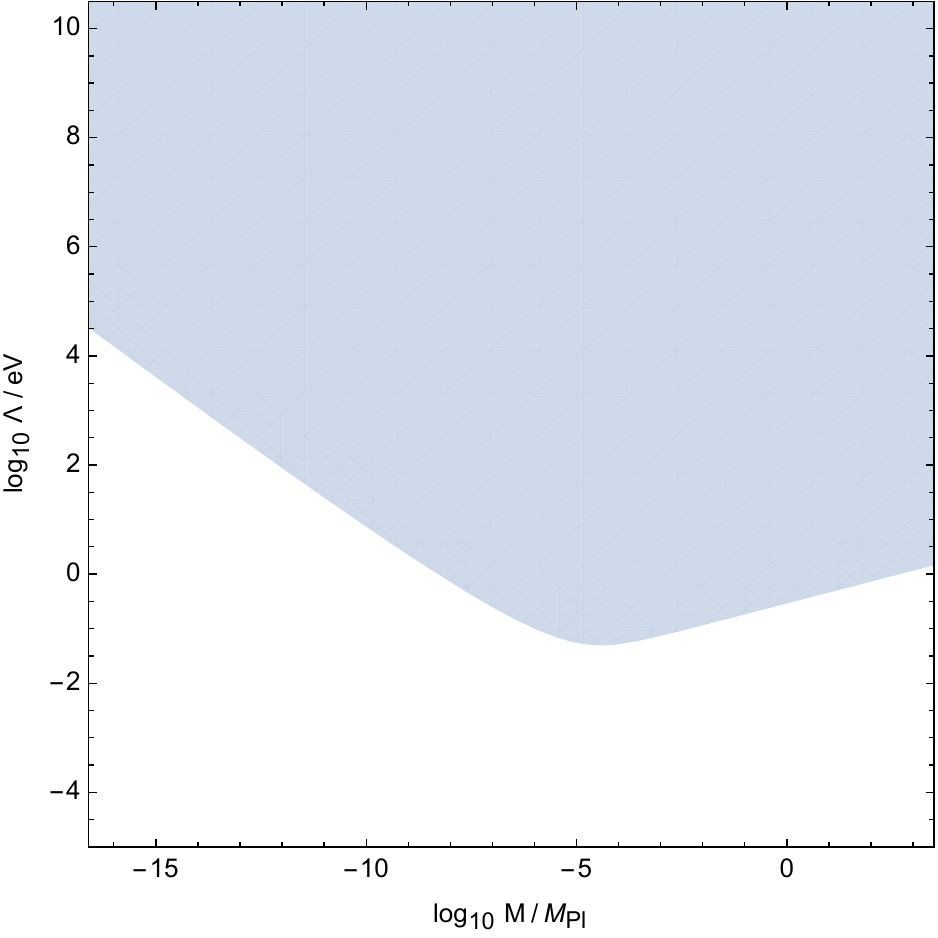}
    \includegraphics[width=0.45\linewidth]{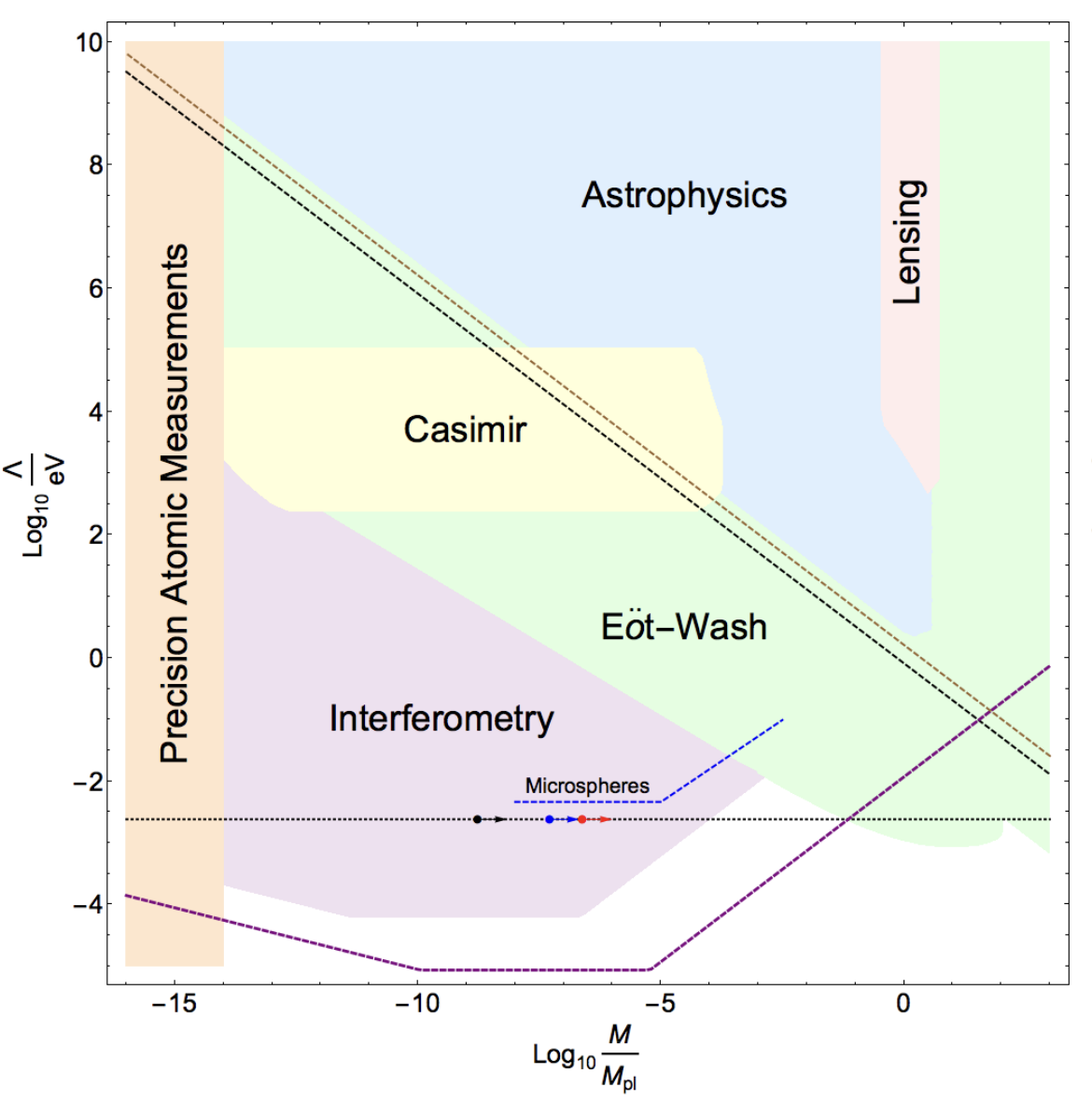}
    \caption{Projected clock bound on chameleon parameter space (left).  On the right are shown existing bounds on chameleon parameter space, reproduced from \cite{Burrage:2016bwy}.  It is seen that all parameter space accessible to the Sun-atomic clocks interaction is ruled out by existing experiments.}
    \label{fig:chameleon-bounds}
\end{figure}

The chameleon scalar charge of an object is
\begin{equation}
    Q_\mathrm{obj} = m_\mathrm{obj} \lambda_\mathrm{obj}~,
\end{equation}
where $\lambda_\mathrm{obj}$ is the ``screening factor''
\begin{equation}
    \lambda_\mathrm{obj} \approx \min \left( \frac{3 M \phi_\mathrm{amb}}{\rho_\mathrm{nucleon} R_\mathrm{obj}^2}, 1 \right)~.
\end{equation}
The scalar field $\phi_\mathrm{amb}$ is the ambient field value in the absence of the object.  We shall assume that $\phi_\mathrm{amb}$ minimizes the effective potential in the Milky Way:
\begin{equation}
    \phi_\mathrm{amb} = \sqrt{\frac{M \Lambda^5}{\rho_\mathrm{MW}}}~,
\end{equation}
where $\rho_\mathrm{MW} \approx m_\odot / \mathrm{pc}^3$ is a typical average galactic density.  The field profile sourced by the Sun may then be solved for by expanding the Lagrangian to quadratic order around $\phi_\mathrm{amb}$.  The solution is then
\begin{equation}
    \phi' = \frac{\lambda_\mathrm{sun} m_\odot}{4 \pi M r^2} e^{- m_\mathrm{eff} r}~,
\end{equation}
where $m_\mathrm{eff}$ is the mass of the scalar fluctuations around $\phi_\mathrm{amb}$:
\begin{equation}
    m_\mathrm{eff}^2 = \frac{2 \Lambda^5}{\phi_\mathrm{amb}^3}~.
\end{equation}
As a first approximation, we ignore the effects of the Earth, as well as the surrounding apparatus and atmosphere around the atomic clocks.  We will also assume the atoms are completely unscreened.  All of these effects would further suppress the Sun-clock interactions.  But even without those effects, the projected constraint is in tension with existing bounds, as seen in Fig.~\ref{fig:chameleon-bounds}.

\section{Coupling to fundamental particles}
\label{app:DM-couplings}

In this work we have described a coupling between the scalar particle and protons and neutrons.  Of course, these couplings are to be understood as effective couplings to these composite particles, arising from couplings to fundamental Standard Model fields, particularly quarks and gluons.  Some works~\cite{Hees:2016gop, Sherrill_2023} use these fundamental couplings, so it is useful to provide a translation between couplings to fundamental fields and to composite particles.  This topic was discussed in detail in~\cite{Damour:2010rp}, and we will briefly summarise the relevant parts of their derivation for this work.  (The reader may also be interested in~\cite{SevillanoMunoz:2024ayh}, which examined this topic in the context of theories that contain a screening mechanism.) 

That work described the fundamental couplings as
\begin{equation}
    {\cal L}_\mathrm{int} = \kappa \phi \left( \frac{d_e}{4 e^2} F^2 - \frac{d_g \beta_3}{2 g_3} F_A^2 - \sum_{j = e,u,d} \left(d_{m_j} + \gamma_{m_i} d_g \right) m_i \bar \psi_i \psi_i \right)~,
\end{equation}
where $\kappa = \sqrt{4 \pi G_\mathrm{N}}$, $F, F_A$ are the electromagnetic and gluon field strength tensors respectively, and the $\psi_i$ are the Standard Model fermion fields.  Here $g_3$ is the QCD coupling and $\beta_3(g_3)$ and $\gamma_m(g_3)$ give the running of the QCD coupling and particle masses, respectively.  These particular choices of couplings were made in order to ensure that the couplings and fermion masses are invariant under the renormalization group.

Our present aim is to provide an approximate relation between the fundamental couplings$d_i$ and the couplings $M, M_e$ used in this work.  This may be done by computing the scalar charge of a given mass $m_A$, which is $Q_A = m_A \alpha_A$ where $\alpha_A$ is~\cite{Damour:2010rp}
\begin{equation}
    \alpha_A = \frac{1}{\kappa m_A} \frac{\partial m_A(\phi)}{\partial \phi}~.
\end{equation}
From Eq.~\eqref{interaction-lagrangian} it follows that
\begin{equation}
    \alpha_{p,n,e} = \frac{\sqrt 2 \Mpl}{M_{p,n,e}}~,
    \label{alpha-big-M}
\end{equation}
for a proton, neutron, and electron, respectively.

We will now compute the $\alpha_i$ in terms of the fundamental couplings.  The reader is referred to \cite{Damour:2010rp} for a more complete description of this calculation, while here we summarise only the essential steps.  First, it was shown that
\begin{equation}
    \alpha_A = d_g + \frac{1}{m_A} \left(
    (d_{\hat m} - d_g) \hat m \frac{\partial m_A}{\partial \hat m} + 
    (d_{\delta m} - d_g) \delta m \frac{\partial m_A}{\partial \delta m} + 
    (d_{m_e} - d_g) m_e \frac{\partial m_A}{\partial m_e} +
    d_e \alpha \frac{\partial m_A}{\partial \alpha} \right)~.
\end{equation}
This is phrased in terms of the average light quark masses and their difference:
\begin{align} 
    \hat m = \frac{1}{2} (m_d + m_u)~, \quad
    \delta m = (m_d - m_u)~, 
\end{align}
as well as their corresponding coupling strengths
\begin{align} \nonumber
d_{\hat m} &= \frac{d_{m_d}m_d + d_{m_u}m_u}{m_d+m_u}~, \\
d_{\delta m} &= \frac{d_{m_d}m_d - d_{m_u}m_u}{m_d - m_u}~.
\end{align}
For a single electron $m_A = m_e$ and we simply have
\begin{equation}
    \alpha_e = d_e~.
\end{equation}
From Eq.~\eqref{alpha-big-M} it immediately follows that
\begin{equation}
    {\frac{\sqrt 2 \Mpl}{M_e} = d_e}~.
\end{equation}
The proton and neutron masses are given as
\begin{align} \nonumber
    m_p &= m_{N3} + \sigma - \frac{1}{2} \delta + C_p \alpha~, \\
    m_n &= m_{N3} + \sigma + \frac{1}{2} \delta + C_n \alpha~.
\end{align}
The first term does not depend on $\hat m$ or $\delta m$ and is therefore irrelevant to the present discussion.  The $C_{p, n}$ describe the electromagnetic binding energy contribution to the nucleon mass.  In addition to being small, in this work, we do not consider any coupling between the scalar and photons, so $d_e = 0$, and these terms may be neglected.  The remaining two terms are 
\begin{align} \nonumber
    \sigma &= \bra{n} \hat m (\bar d d - \bar u u) \ket{n} \approx 45~\mathrm{MeV}~, \\
    \delta &= \bra{n} \delta m (\bar d d - \bar u u) \ket{n} \approx 3.1~\mathrm{MeV}~.
\end{align}
With these definitions, the scalar charge of the proton becomes
\begin{align} \nonumber
    \alpha_p &= d_g + (d_{\hat m} - d_g) \frac{\sigma}{m_p} - \frac{1}{2} (d_{\delta m} - d_g) \frac{\delta}{m_p}~,
\end{align}
and the scalar charge of the neutron becomes
\begin{align} \nonumber
    \alpha_n &= d_g + (d_{\hat m} - d_g) \frac{\sigma}{m_n} + \frac{1}{2} (d_{\delta m} - d_g) \frac{\delta}{m_n}~. 
\end{align}
From these results, we see that the gluon binding energy dominates the composition of the nucleons, as expected.
Summarizing these results and substituting numerical values, we have the following mapping between couplings:
\begin{align} \nonumber
    \frac{\sqrt 2 \Mpl}{M_p} &= 0.954 d_g + 0.0168 d_{m_u} + 0.0296 d_{m_d}~, \\ \nonumber
    \frac{\sqrt 2 \Mpl}{M_n} &= 0.950 d_g + 0.0138 d_{m_u} + 0.0357 d_{m_d}~, \\
    \frac{\sqrt 2 \Mpl}{M_e} &= d_{m_e}~.
\end{align}

Having obtained couplings to protons and neutrons, we are in a position to test the procedure that was used to compute scalar charges in the main text.  In Appendix~\ref{sec:theory}, the scalar charge was computed as a weighted average of the various components.  The analogous procedure here for an atom is
\begin{equation}
    \alpha_\mathrm{atom} = \frac{1}{m_\mathrm{atom}} \left( Z m_p \alpha_p + (A - Z) m_n \alpha_n + Z m_e \alpha_e \right)~,
\end{equation}
for an atom of atomic number $Z$, mass number $A$, and total mass $m_\mathrm{atom} = Z m_p + (A-Z) m_n + Z m_e$.  Using titanium-48 as an example, we find
\begin{equation}
    \alpha_\mathrm{Ti} = 0.952 d_g + 0.0152 d_u + 0.0329 d_d + 2.49\times 10^{-4} d_{m_e}~.
\end{equation}
This result, like the expressions in the main text, does not account for the strong force binding energy between nucleons.  This is accounted for in a general expression for atoms that is given in~\cite{Damour:2010rp}.  Using their general result, we find
\begin{equation}
    \alpha_\mathrm{Ti} = 0.917 d_g + 0.0262 d_u + 0.0565 d_d + 2.52\times 10^{-4} d_{m_e}~.
\end{equation}
We see that accounting for the binding energy between nucleons changes some of the coefficients slightly but does not alter the overall picture: $\sim 90\%$ of the coupling is dominated by the scalar-gluon coupling, the quarks' rest mass contributes at the level of a few per cent, and the electron rest mass is sub-leading.



\newpage
\twocolumngrid
\nocite{*}
\bibliography{aipsamp}

\end{document}